\documentclass[a4paper,11pt]{article}
\pdfoutput=1
\usepackage{jheppub} 
\usepackage[T1]{fontenc} % if needed
\usepackage{graphicx,xcolor}
\usepackage{bbold}
\usepackage{float}
\usepackage{mathtools,braket,blkarray}
\allowdisplaybreaks
\usepackage{amssymb,amsfonts,bbm,relsize}
\usepackage[width=0.9\textwidth,footnotesize]{caption}
\usepackage[font=scriptsize]{subcaption}
\hypersetup{pageanchor=false}
\usepackage{booktabs,multirow,makecell}
\usepackage[utf8]{inputenc}
\usepackage{cleveref}

\newcommand{\pmatr}[1]{\begin{pmatrix} #1 \end{pmatrix}}

\newcommand{\RK}{$R_{K^{(*)}} \text{ } $}
\newcommand{\be}{\begin{equation}}
\newcommand{\ee}{\end{equation}}
\newcommand{\overbar}[1]{\mkern 1.5mu\overline{\mkern-1.5mu#1\mkern-1.5mu}\mkern 1.5mu}

\title{\boldmath Flavourful $Z'$ portal for vector-like neutrino Dark Matter and $R_{K^{(*)}}$}

\author[a]{Adam Falkowski,}
\author[b]{Stephen F. King,}
\author[b]{Elena Perdomo}
\author[a]{and Mathias Pierre}

\affiliation[a]{Laboratoire de Physique Th\'eorique, CNRS, Univ. Paris-Sud, Universit\'e Paris-Saclay, \\91405 Orsay, France}
\affiliation[b]{School of Physics \& Astronomy, University of Southampton, \\Southampton SO17 1BJ, UK}

\emailAdd{adam.falkowski@th.u-psud.fr}
\emailAdd{S.F.King@soton.ac.uk}
\emailAdd{e.perdomo-mendez@soton.ac.uk}
\emailAdd{mathias.pierre@th.u-psud.fr}

\abstract{
We discuss a flavourful $Z'$ portal model with a coupling to fourth-family singlet Dirac neutrino dark matter.  
In the absence of mixing, the $Z'$ is fermiophobic, having no couplings to the three chiral families, but does couple to a fourth vector-like family. 
Due to mixing effects, the $Z'$ gets induced couplings to second family left-handed lepton doublets and third family left-handed quark doublets. 
This model can simultaneously account for the measured $B$-decay ratios $R_{K}$ and $R_{K^*}$ and for the observed relic abundance of dark matter. 
We identify the parameter space where this explanation is consistent with existing experimental constraints from dark matter direct and indirect detection, LHC searches, and precision measurements of flavour mixing and neutrino processes.  
}

\begin{document}

\preprint{LPT-Orsay-18-16, Southampton-18-16}

\maketitle
\flushbottom

\section{Introduction}
\label{sec:intro}

Recently, the phenomenological motivation for considering non-universal $Z'$ models has increased due to  mounting evidence for semi-leptonic $B$ decays whose rates and differential distributions are inconsistent with those predicted by the Standard Model (SM)~\cite{Descotes-Genon:2013wba,Altmannshofer:2013foa,Ghosh:2014awa}.
In particular, the LHCb Collaboration has reported a number of deviations from $\mu$-$e$ universality  in  $B\rightarrow K^{(*)}l^+l^-$  decays. 
The ratios of $\mu^+ \mu^-$ to $e^+ e^-$ final states:  $R_K$~\cite{Aaij:2014ora} and $R_{K^*}$~\cite{Aaij:2017vbb} are observed to be about $70\%$ of their expected values, each displaying a $2.5\sigma$ deviation  from the SM.  
Combining that with the input from other  $b\rightarrow s \ell^+ \ell^-$ processes, the SM is disfavored by $4$ to $5$ standard deviations~\cite{Capdevila:2017bsm,Altmannshofer:2017yso}. 
  
The $R_K$ and $R_{K^*}$ anomalies could be the first evidence of new physics. 
A number of recent phenomenological analyses, 
see e.g. \cite{Capdevila:2017bsm,Altmannshofer:2017yso,Ciuchini:2017mik,Hiller:2017bzc,Geng:2017svp,Ghosh:2017ber,Bardhan:2017xcc,DAmico:2017mtc,Alok:2017sui}, conclude that these data can be well fit  
when the low-energy Lagrangian below the weak scale contains a new physics operator of the $C^\text{NP}_{9\mu}=-C^\text{NP}_{10\mu}$ form:  
\begin{equation}
\Delta \mathcal{L}_\text{eff} \supset 
G_{b s\mu} (\bar{b}_L \gamma^\mu s_L) (\bar{\mu}_L \gamma_\mu \mu_L) + {\rm h.c.}, 
\qquad G_{bs\mu} \sim \frac{1}{(30\text{ TeV})^2}. 
\label{eq:c9mc10}
\end{equation}

In a flavourful $Z'$ model, the new physics operator in Eq.~\ref{eq:c9mc10} will arise from tree-level $Z'$ exchange: $G_{b s\mu} = -\frac{g_{bs} g_{\mu\mu}}{M_{Z^\prime}^2}$, where 
$g_{bs}$ is the flavour-violating $Z'$ coupling to left-handed b- and s-quarks, and $g_{\mu \mu}$ is the couplings to left-handed muons.  
There is already a vast literature discussing the $Z'$ explanation of the B-anomalies and phenomenological constraints on the parameter space of such models, see e.g.~\cite{Gauld:2013qba,Buras:2013qja,Altmannshofer:2014cfa,Crivellin:2015mga,Crivellin:2015lwa,Niehoff:2015bfa,Celis:2015ara,Greljo:2015mma,Niehoff:2015iaa,Altmannshofer:2015mqa,Falkowski:2015zwa,Carmona:2015ena,GarciaGarcia:2016nvr,Megias:2016bde,Chiang:2016qov,Altmannshofer:2016oaq,Boucenna:2016qad,Foldenauer:2016rpi,Kamenik:2017tnu,Chivukula:2017qsi,Faisel:2017glo,Ellis:2017nrp,Alonso:2017uky,Carmona:2017fsn,Dalchenko:2017shg,Raby:2017igl,Bian:2017rpg,Bian:2017xzg,Alok:2017jgr,Fox:2018ldq,Chala:2018igk}.
In realistic models of this kind, the coupling  $g_{bs}$ is strongly constrained by precision measurements of the $B_s$ meson mass difference.
Taking that into account, one can derive the constraint $M_{Z'} \lesssim 1.2 g_{\mu \mu}$~TeV, implying that $M_{Z'}$ must be close to the weak scale in weakly coupled models. 
The corollary is that the $Z'$ is in the correct mass range to act as mediator between the SM and  thermally produced dark matter~\cite{Sierra:2015fma,Belanger:2015nma,Celis:2016ayl,Altmannshofer:2016jzy,Baek:2017sew,Cline:2017qqu,Fuyuto:2017sys}. 
In this paper we further pursue this direction, and discuss a $Z'$ model that can account for the B-anomalies and, simultaneously, explain the observed relic abundance via a weakly interacting massive particle (WIMP) communicating with the SM through the same $Z'$.

We follow Ref.~\cite{King:2017anf},  
which introduces a fourth vector-like family with non-universal gauged $U(1)'$ charges.
The idea is that the $Z'$ couples universally to the three chiral families, which then mix with the non-universal fourth family to induce effective non-universal couplings in the physical light mixed quarks and leptons. 
Such a mechanism has wide applicability, for example it was recently discussed in the context of 
F-theory models with non-universal gauginos \cite{Romao:2017qnu}.
Two explicit examples were discussed in \cite{King:2017anf}. 
Firstly an $SO(10)\rightarrow SU(5)\times U(1)_X$  model, where we identified $U(1)'\equiv U(1)_X$,
which however was subsequently shown to be not consistent with both explaining $R_{K^*}$ and respecting the $B_s$ mass difference \cite{Antusch:2017tud}.
Ref.~\cite{King:2017anf} also  discussed a fermiophobic model where the gauged $U(1)'$ charges are not carried by the three chiral families, only by fourth vector-like family.
In the absence of mixing, the $Z'$ is fermiophobic, having no couplings to the three chiral families, but does couple to a fourth vector-like family. 
Due to mixing effects, we shall suppose that the $Z'$ gets induced couplings to second family left-handed lepton doublets (containing the left-handed muon and its neutrino) and third family left-handed quark doublets (containing the left-handed top and bottom quarks). 
Including only such couplings is enough to address the B-anomalies, in analogy to related scenarios where new  vector-like fermions mix with the SM ones~\cite{Niehoff:2015bfa,Niehoff:2015iaa,Carmona:2015ena,Boucenna:2016qad,Megias:2016bde,Carmona:2017fsn,Raby:2017igl,Chala:2018igk}. 
In addition, this set-up  provides a natural WIMP dark matter candidate: the neutrino residing in the fourth family. 
We  are interested in the parameter space of this model where both B-anomalies and the relic abundance of dark matter are simultaneously explained. 
We show that this can be achieved without conflicting a myriad of direct and indirect dark matter constraints as well as experimental constraints such as $B_s$ mixing, LHC searches, neutrino trident, and so on. 
The requirement to satisfy all these constraints in a natural way  points to a specific corner of the parameter space, with $300~{\rm GeV} \lesssim m_{Z'} \lesssim 1$~TeV, dark matter heavier than a TeV, and a narrow range of possible $Z'$ couplings. 

This paper is organized as follows. 
In section~\ref{sec:model} we define our gauged $U(1)'$ model with a vector-like fourth family.
The $Z'$ couplings relevant for the subsequent analysis are summarized in Eq.~\ref{eq:Zp_Rk_couplings} . 
In section~\ref{constraints} we discuss the constraints these parameters need to satisfy in order to address the B-anomalies without conflicting other experimental results. 
In section~\ref{darkmatter} we turn to the dark matter sector, and identify the masses and  couplings of the vector-like fourth family singlet Dirac neutrino which lead to a correct relic density, while evading all indirect and direct searches so far. 
Our main results are contained in Section~\ref{conclusion}, where we put together the requirements imposed by the B-anomalies and by the relic density, and identify the viable parameter space where both are satisfied.  

%%%%%%%%%%%%%%%%%%%%%%%%%%%%%%%%%%%%%%%%%%%%%%%%%%%%%%%%%%%%%%%%%%%%%%%%
\section{The model}
\label{sec:model}
We consider a model in which, in addition to the SM with the usual three chiral families of left-handed quarks and leptons, including the right-handed neutrinos, we add a dark $U(1)^\prime$ gauge symmetry and a fourth vector-like family of fermions. The idea is to have the SM quarks and leptons neutral under the  $U(1)^\prime$ while the vector-like family has the SM quantum numbers and is charged under the  $U(1)^\prime$, leading to a dark matter candidate and flavour-changing $Z^\prime$ operators after the vector-like fermion mass term mix with the SM fermions. 

Table~\ref{tab:model} shows all the particle content and their corresponding representations and charges. The non-universal $U(1)^\prime$ charges forbid mixing between the fourth family and the chiral families via the usual Higgs Yukawa couplings. Therefore, we need to add new singlet scalars, with appropriate $U(1)^\prime$ charges, to generate mass mixing of quarks and leptons with the vector-like family. The $U(1)^\prime$ is broken by the VEVs of the new Higgs singlets $\phi_\psi$ to yield a massive $Z^\prime$.
\begin{table}
		\centering
		\begin{tabular}[t]{| c | c c c c|}
			\hline
			\multirow{2}{*}{\rule{0pt}{4ex}Field}	& \multicolumn{4}{c |}{Representation/charge} \\
			\cline{2-5}
			\rule{0pt}{3ex}			& $SU(3)_c$ & $SU(2)_L$ &  $ U(1)_Y$ & $ U(1)^\prime $  \\ [0.75ex]
			\hline \hline
			\rule{0pt}{3ex}%
			$Q_{Li}$ & $\bf{3}$ &$\bf{2}$   &$1/6$ & $ 0 $ \\
			$u_{Ri}$ & $\bf{3}$ &$\bf{1}$   &$2/3$ & $ 0 $ \\
			$d_{Ri}$ & $\bf{3}$ &$\bf{1}$   &$-1/3$\phantom{+} & $ 0 $ \\
			$L_{Li}$ & $\bf{1}$ &$\bf{2}$   &$-1/2$\phantom{+} & $ 0 $ \\
			$e_{Ri}$ & $\bf{1}$ &$\bf{1}$   &$-1$\phantom{+} & $ 0 $ \\
			$\nu_{Ri}$ & $\bf{1}$ &$\bf{1}$   &$0$ & $ 0 $ \\
			\hline
			\hline
			$H$ & $\bf{1}$ &$\bf{2}$   &$1/2$ & $ 0 $ \\
			\hline
			\hline 
			$Q_{L4},\tilde{Q}_{R4} $ & $\bf{3}$ &$\bf{2}$   &$1/6$ & $ q_{Q4} $ \\
			$u_{R4}, \tilde{u}_{L4}$ & $\bf{3}$ &$\bf{1}$   &$2/3$ & $ q_{u4} $ \\
			$d_{R4}, \tilde{d}_{L4}$ & $\bf{3}$ &$\bf{1}$   &$-1/3$\phantom{+} & $ q_{d4} $ \\
			$L_{L4}, \tilde{L}_{R4}$ & $\bf{1}$ &$\bf{2}$   &$-1/2$ \phantom{+}& $ q_{L4} $ \\
			$e_{R4}, \tilde{e}_{L4}$ & $\bf{1}$ &$\bf{1}$   &$-1$\phantom{+} & $ q_{e4} $ \\
			$\nu_{R4}, \tilde{\nu}_{L4}$ & $\bf{1}$ &$\bf{1}$   &$0$ & $ q_{\nu4} $ \\
			\hline
			\hline
				$\phi_{Q,u,d,L,e}$ & $\bf{1}$ &$\bf{1}$   &$0$ & $ -q_{Q_4,u_4,d_4,L_4,e_4} $ \\
				\hline
		\end{tabular}
		\caption{The model consists of the usual three chiral families of quarks and leptons $\psi_i$ $(i=1,2,3)$, including the right-handed neutrino, a Higgs doublet $H$, plus a fourth vector-like family of fermions $\psi_4, \tilde{\psi}_4$ and new Higgs singlets $\phi_\psi$ which mix fourth family fermions with the three chiral families. Note that we exclude $\phi_\nu$ so that $\nu_{R4}, \tilde{\nu}_{L4}$ do not mix and are stable.}
		\label{tab:model}
	\end{table}

The Higgs Yukawa couplings of the first three chiral families can be written in a $4\times 4$ matrix notation
\begin{equation}
\mathcal{L}^\text{Yukawa}= y^u \bar{Q}_{L} \tilde{H} u_{R}+ y^d \bar{Q}_{L} H  d_{R}+y^e  \bar{L}_{L} H e_{R}  +y^\nu \bar{L}_{L}  \tilde{H} \nu_{R} +\text{h.c.}~,
\label{eq:yuk1}
\end{equation}
where $\tilde{H}=i\sigma_2 H^*$ and $y^u,$ $y^d$, $y^e$, $y^\nu$ are $4\times 4 $ matrices with the fourth row and columns consisting of all zeros, since the fourth family does not couple to the Higgs doublets. The $U(1)^\prime$ charges allow Yukawa couplings between the singlet fields $\phi$, the fourth family $\tilde{\psi}_4 $ and the first three chiral families $\psi_i$. Furthermore, there is an explicit mass term between the opposite chirality fourth family fields $\psi_4$ and $\tilde{\psi}_4$,
\begin{equation}
\begin{split}
\mathcal{L}^\text{mass}&=x_i^Q \phi_Q \bar{Q}_{Li}\tilde{Q}_{R4}+x_i^u \phi_u \bar{\tilde{u}}_{L4}u_{Ri}+x_i^d \phi_d \bar{\tilde{d}}_{L4}d_{Ri}+x_i^L \phi_L \bar{L}_{Li}\tilde{L}_{R4}+x_i^e \phi_e \bar{\tilde{e}}_{L4}e_{Ri} \\
 &+ M_4^Q \bar{Q}_{L4}\tilde{Q}_{R4}+ M_4^u \bar{\tilde{u}}_{L4}u_{R4} + M_4^d \bar{\tilde{d}}_{L4}d_{R4} +M_4^L \bar{L}_{L4}\tilde{L}_{R4}+M_4^e \bar{\tilde{e}}_{L4}e_{R4} \\
 &+M_4^\nu \bar{\tilde{\nu}}_{L4}\nu_{R4} 
+\text{h.c.}~, 
\end{split}
\end{equation}
where $i=1,...,3$. 

The fourth-family vector-like singlet neutrinos 
$\nu_{R4},\tilde{\nu}_{L4}$ 
are special since we don't have a singlet field $\phi_\nu$ that couples them to the other families, which is why such terms are absent in the above equation. This implies that $\nu_{R4},\tilde{\nu}_{L4}$ are absolutely stable, with their stability guaranteed by an unbroken global $U(1)_{\nu_{R4}}$ and, since they do not carry any Standard Model quantum numbers, they may play the role of dark matter.  Note that we also impose lepton number conservation $U(1)_L$ for all four families of leptons which forbids Majorana mass terms. Hence
all neutrinos (including those in the fourth vector-like family) will have purely Dirac masses.\footnote{Alternatively it is possible to introduce various seesaw mechanisms into this kind of model, leading to Majorana masses, as recently discussed \cite{Antusch:2017tud}. However in this paper we only consider Dirac neutrinos.}

After the singlet scalar fields $\phi$ obtain a non-zero vacuum expectation value (VEV), we may rewrite the Lagrangian in terms of new mass parameters $M_i^Q=x_i^Q\left<\phi_Q\right>$, similarly for the other mass parameters, such that
\begin{equation}
\begin{split}
\mathcal{L}^\text{mass}&=M_\alpha^Q \bar{Q}_{L\alpha}\tilde{Q}_{R4}+ M_\alpha^u \bar{\tilde{u}}_{L4}u_{R\alpha} + M_\alpha^d \bar{\tilde{d}}_{L4}d_{R\alpha} +M_\alpha^L \bar{L}_{L\alpha}\tilde{L}_{R4}+M_\alpha^e \bar{\tilde{e}}_{L4}e_{R\alpha} \\
 &+M_4^\nu \bar{\tilde{\nu}}_{L4}\nu_{R4} 
+\text{h.c.}~, 
\end{split}
\end{equation}
where $\alpha=1,...,4$. We may diagonalize the mass matrix before electroweak symmetry breaking, when only the fourth family is massive 
\begin{equation}
\begin{split}
\mathcal{L}^\text{mass}&=\tilde{M}_4^Q \bar{Q'}_{L4}\tilde{Q}_{R4}+ \tilde{M}_4^u \bar{\tilde{u}}_{L4}u'_{R4} + \tilde{M}_4^d \bar{\tilde{d}}_{L4}d'_{R4} +\tilde{M}_4^L \bar{L'}_{L4}\tilde{L}_{R4}+\tilde{M}_4^e \bar{\tilde{e}}_{L4}e'_{R4} \\
 &+M_4^\nu \bar{\tilde{\nu}}_{L4}\nu_{R4} 
+\text{h.c.}
\end{split}
\end{equation}
The prime states for the heavy mass basis where only the fourth family has explicit vector-like Dirac mass terms and it's related to the original charge basis by unitary mixing matrices, 
\begin{equation}
Q'_L=V_{Q_L}Q_L, \quad u'_R=V_{u_R}u_R, \quad d'_R=V_{d_R}d_R, \quad L'_L=V_{L_L}L_L, \quad e'_R=V_{e_R}e_R,
\label{eq:transf_heavybasis}
\end{equation}
while for the neutrino states $\tilde{\nu}_{L4}$ and $\nu_{R4}$ the original and the mass basis coincides. In this basis, the Yukawa couplings in Eq.~\ref{eq:yuk1} become
\begin{equation}
\mathcal{L}^\text{Yukawa}= y'^u \bar{Q'}_{L} \tilde{H} u'_{R}+  y'^d \bar{Q'}_{L} H d'_{R}+y'^e \bar{L'}_{L} H  e'_{R} + y'^\nu \bar{L'}_{L} \tilde{H} \nu_{R}+ \text{h.c.}~, 
\label{eq:yuk2}
\end{equation}
where
\begin{equation}
y'^u=V_{Q_L}y^u V^\dagger_{u_R}, \quad y'^d=V_{Q_L}y^d V^\dagger_{d_R}, \quad y'^e=V_{L_L}y^e V^\dagger_{e_R} \quad y'^\nu=V_{L_L}y^\nu.
\end{equation}
This shows that there is a coupling between the heavy fourth family and the Higgs due to their mixing with the first three chiral families. However, this coupling will be small since the original  $y^u,$ $y^d$, $y^e$, $y^\nu$ contain zeroes in the fourth row and column and they are mixing suppressed. Therefore, we can integrate out the fourth family and look at the low energy effective theory by simply removing the fourth rows and columns of the primed Yukawa matrices in Eq.~\ref{eq:yuk2}. The three massless families, below the heavy mass scale, are described by 
\begin{equation}
\mathcal{L}^\text{Yukawa}_\text{light}= y'^u_{ij} \bar{Q'}_{Li} \tilde{H} u'_{Rj}+ y'^d_{ij} \bar{Q'}_{Li}H d'_{Rj}+y'^e_{ij} \bar{L'}_{Li} H  e'_{Rj}+y'^\nu_{ij} \bar{L'}_{Li} \tilde{H}  \nu_{Rj} + \text{h.c.}~, 
\label{eq:yuk_light}
\end{equation}
where 
\begin{equation}
y'^u_{ij}=(V_{Q_L}y^u V^\dagger_{u_R})_{ij}, \quad y'^d_{ij}=(V_{Q_L}y^d V^\dagger_{d_R})_{ij}, \quad y'^e_{ij}=(V_{L_L}y^e V^\dagger_{e_R})_{ij}, \quad y'^\nu_{ij}=(V_{L_L}y^\nu )_{ij}
\end{equation}
and $i,j=1,...,3$. The Yukawa matrices for the quarks and charged leptons can be now diagonalized 
\begin{equation}
V'_{uL}y'^uV'^\dagger_{uR}=\text{diag}(y_u,y_c, y_t), \quad V'_{dL}y'^d V'^\dagger_{dR}=\text{diag}(y_d,y_s,y_b), \quad V'_{eL}y'^e V'^\dagger_{eR}=\text{diag}(y_e,y_\mu,y_\tau).
\label{eq:diag_Yuk_mat}
\end{equation}
The unitary CKM matrix is then given by
\begin{equation}
V_\text{CKM}=V'_{uL}V'^\dagger_{dL}.
\end{equation}
In the case of neutrinos, since we are forbidding Majorana masses, the light physical neutrinos have Dirac mass eigenvalues given by,
\begin{equation}
v V'_{\nu L}y'^\nu V'^\dagger_{\nu R}=\text{diag}(m_1, m_2, m_3).
\label{eq:nu_diag}
\end{equation}
The lepton mixing matrix or PMNS matrix can be constructed from the transformations in eqs.~\ref{eq:diag_Yuk_mat} and~\ref{eq:nu_diag}
\begin{equation}
V_\text{PMNS}=V'_{eL} V'^\dagger_{\nu L}.
\end{equation}
To look at the Lagrangian involving the SM gauge couplings, we emphasize that all the four families have the same charges under the SM. The unitary transformations in Eq.~\ref{eq:transf_heavybasis} cancel as in the usual GIM mechanism and the gauge couplings in the heavy mass basis remains the same as in the SM. After integrating out the fourth family and electroweak symmetry is broken, and the light Yukawa matrices are diagonalised, the couplings to the $W^\pm$ gauge bosons are 
\begin{equation}
\begin{split}
\mathcal{L}^\text{int}_W&=\frac{g_2}{\sqrt{2}}\pmatr{\bar{u}_L  & \bar{c}_L  &\bar{t}_L }V_\text{CKM}W^+_\mu \gamma^\mu \pmatr{d_L \\ s_L \\ b_L}  \\
&+ \frac{g_2}{\sqrt{2}}\pmatr{\bar{e}_L  & \bar{\mu}_L  &\bar{\tau}_L }V_\text{PMNS}W^+_\mu \gamma^\mu \pmatr{\nu_{1L} \\ \nu_{2L} \\ \nu_{3L}} + \text{h.c.},
\end{split}
\end{equation}
where $g_2$ is the usual $SU(2)_L$ gauge coupling. For the couplings to the $Z$ gauge boson, the same happens, the charges are the same for the fourth families and the transformations in Eq.~\ref{eq:transf_heavybasis} cancel, such that in the heavy mass basis, after electroweak symmetry breaking, we are left with
\begin{equation}
\mathcal{L}^\text{int}_Z=\frac{e}{2 s_W c_W}\bar{\psi}'_\alpha Z_\mu \gamma^\mu (C^\psi_V-C^\psi_A \gamma_5) \psi'_\alpha
\label{eq:Z_couplings}
\end{equation}
where
\begin{equation}
\psi'_\alpha=u'_\alpha, d'_\alpha, e'_\alpha, \nu'_\alpha \quad \alpha=1,...,4
\end{equation}
and
\begin{equation}
C^\psi_A=t_3, \quad C^\psi_V=t_3-2s^2_W Q.
\end{equation}
The electric charge of the fermions is denoted by $Q$ and $t_3$ are the eigenvalues of $\sigma_3/2$. The couplings to the $Z$ boson are flavour diagonal, even after diagonalization of the light fermion mass matrices, due to the unitary transformations cancelling. The interactions will be the same as in Eq.~\ref{eq:Z_couplings}, replacing the fields $\psi'_\alpha$ by their three family mass eigenstates. 

In the case of the couplings to the $Z'$ gauge bosons, we have non-universal couplings that lead to flavour changing. In the original basis, after the $U(1)'$ symmetry is broken, we have diagonal gauge couplings between the massive $Z'$ gauge boson and the four families
\begin{equation}
\mathcal{L}^\text{gauge}_{Z'}=g' Z^\prime_\mu (\bar{Q}_L D_Q \gamma^\mu Q_L+\bar{u}_R D_u \gamma^\mu u_R+\bar{d}_R D_d \gamma^\mu d_R 
+  \bar{L}_L D_L \gamma^\mu L_L+\bar{e}_R D_e \gamma^\mu e_R)
\end{equation}
where, 
\begin{equation}
\begin{split}
D_Q=\text{diag}(0,0,0,q_{Q4}), \quad D_u=\text{diag}(0,0,0,q_{u4}), \quad D_d=\text{diag}(0,0,0,q_{d4}) \\
D_L=\text{diag}(0,0,0,q_{L4}), \quad D_e=\text{diag}(0,0,0,q_{e4}), \quad D_\nu=\text{diag}(0,0,0,q_{d4}).
\end{split}
\end{equation}
In addition there are the fourth family couplings involving the opposite chirality states $\tilde{\psi}_4$. Using the transformations in Eq.~\ref{eq:transf_heavybasis}, we get the $Z'$ couplings in the diagonal heavy mass basis
\begin{equation}
\mathcal{L}^\text{gauge}_{Z'}=g' Z^\prime_\mu (\bar{Q'}_L D'_Q \gamma^\mu Q'_L+\bar{u'}_R D'_u \gamma^\mu u'_R+\bar{d'}_R D'_d \gamma^\mu d'_R 
+  \bar{L'}_L D'_L \gamma^\mu L'_L+\bar{e'}_R D'_e \gamma^\mu e'_R)
\end{equation}
where $D'_Q=V_{Q_L}D_QV^\dagger_{Q_L}$, and similarly with $Q\rightarrow L$, etc. Ignoring phases, these matrices can be parametrized as
\begin{equation}
D'_Q=q_{Q_4} \left(
\begin{array}{cccc}
 s_{14}^2 & c_{14} s_{14} s_{24} & c_{14} c_{24} s_{14} s_{34} & c_{14} c_{24} c_{34} s_{14} \\
 c_{14} s_{14} s_{24} & c_{14}^2 s_{24}^2 & c_{14}^2 c_{24} s_{24} s_{34} & c_{14}^2 c_{24} c_{34} s_{24} \\
 c_{14} c_{24} s_{14} s_{34} & c_{14}^2 c_{24} s_{24} s_{34} & c_{14}^2 c_{24}^2 s_{34}^2 & c_{14}^2 c_{24}^2 c_{34} s_{34} \\
 c_{14} c_{24} c_{34} s_{14} & c_{14}^2 c_{24} c_{34} s_{24} & c_{14}^2 c_{24}^2 c_{34} s_{34} & c_{14}^2 c_{24}^2 c_{34}^2 \\
\end{array}
\right)
\end{equation}
where $s_{ij}$ and $c_{ij}$ refer to $\sin\theta_{ij}$ and $\cos\theta_{ij}$ (we have also suppressed the superscript in the angles  $ s^Q_{14} \rightarrow  s_{14}$ for simplicity). Since the $U(1)'$ charges differ for the fourth family, the unitary transformations do not cancel and the matrices $D'_Q$, etc., are not generally diagonal. Therefore, $Z'$ exchange can couple to light families of different flavour.

We are interested in the $\bar{s}bZ'$ and $\bar{\mu}\mu Z'$ couplings, needed for the $R_K$ anomaly. Assuming that only the mixing angles $\theta^{Q_L}_{34}$ and $\theta^{L_L}_{24}$ are different from zero\footnote{A more natural possibility would be to assume that the new vector-like fermions have a large mixing only with the 3rd generation of the SM doublet, that is with taus instead of muons. 
Then the coupling to muons could arise due to a mixing between the SM charged leptons, as in \cite{King:2017anf}. 
However,  explaining the B-meson anomalies in such a set-up  runs in conflict with the strong bounds from  non-observation of $\tau \rightarrow 3 \mu$. 
}
the mixing mass matrices become
\begin{equation}
D'_Q=q_{Q_4}\left(
\begin{array}{cccc}
 0 & 0 & 0 & 0 \\
 0 & 0 & 0 & 0 \\
 0 & 0 & (s^Q_{34})^2 & c^Q_{34} s^Q_{34} \\
 0 & 0 & c^Q_{34} s^Q_{34} & (c^Q_{34})^2 \\
\end{array}
\right),
\quad
D'_L=q_{L_4} \left(
\begin{array}{cccc}
 0 & 0 & 0 & 0 \\
 0 & (s^L_{24})^2 & 0 & c^L_{24} s^L_{24} \\
 0 & 0 & 0 & 0 \\
 0 & c^L_{24} s^L_{24} & 0 & (c^L_{24})^2 \\
\end{array}
\right)
\end{equation}
while the rest of them being zero. In the low energy effective theory, after integrating out the fourth heavy family, the $Z'$ couplings to the three massless families of quarks and leptons are
\begin{equation}
\mathcal{L}_{Z'}^\text{gauge}= g'Z'_{\mu} \left(q_{Q_4}(s^Q_{34})^2\bar{Q'}_{L_3} \gamma^\mu Q'_{L_3}+ q_{L_4}(s^L_{24})^2\bar{L'}_{L_2} \gamma^\mu L'_{L_2} \right),
\label{eq:Z'_general_couplings}
\end{equation}
where $Q'_{L3}=(t'_L,b'_L)$ and $L'_{L2}=(\nu'_{\mu L}, \mu'_L)$. Using now the diagonalization of the Yukawa matrices in Eq.~\ref{eq:diag_Yuk_mat}, we can expand the primed fields in terms of the mass eigenstates, 
\begin{eqnarray}
b'_L&=&(V'^\dagger_{dL})_{31} d_L +(V'^\dagger_{dL})_{32}s_L+(V'^\dagger_{dL})_{33}b_L \nonumber \\
t'_L&=&(V'^\dagger_{uL})_{31} u_L +(V'^\dagger_{uL})_{32}c_L+(V'^\dagger_{uL})_{33}t_L \nonumber \\
\nu'_{\mu L}&=&(V'^\dagger_{\nu L})_{21} \nu_{1 L} +(V'^\dagger_{\nu L})_{22}\nu_{2 L}+(V'^\dagger_{\nu L})_{23} \nu_{3 L} \\
\mu'_L&=&(V'^\dagger_{eL})_{21} e_L +(V'^\dagger_{eL})_{22}\mu_L+(V'^\dagger_{eL})_{23}\tau_L. \nonumber
\end{eqnarray}
For simplicity, we assume that the charged lepton mass matrix is diagonal so that we may drop the primes 
on the muon field so that $\mu'_L = \mu_L$. Under this assumption, in the lepton sector,
the $Z'$ only couples to muon mass eigenstates $\mu_L$ and muon neutrinos $\nu_{\mu L}$,
where the latter are related to neutrino mass eigenstates by the PMNS matrix,
\begin{equation}
\nu'_{\mu L} = (V_\text{PMNS})_{21} \nu_{1 L} +(V_\text{PMNS})_{22}\nu_{2 L}+(V_\text{PMNS})_{23} \nu_{3 L} 
\end{equation}
Given the hierarchies of the CKM matrix, we will assume similar hierarchies of the rotation matrix elements:
\begin{equation}
\lvert (V'_{(d,u)L})_{31} \rvert^2 \ll \lvert (V'_{(d,u)L})_{32}  \rvert^2 \ll \lvert(V'_{(d,u)L})_{33}  \rvert^2 \approx 1 
\label{eq:mixing_supp}
\end{equation}

The vector-like neutrino $\nu_4$ is not charged under the SM and it is considered as a dark matter candidate. The portal that allows it to annihilate into ordinary matter is the $Z'$ mediator. 
The explicit coupling between the $Z'$ and the dark matter candidate $\nu_4$ is
\begin{equation}
\mathcal{L}_{Z'}^{\nu_4} = g'q_{\nu_4} Z'_\mu \overbar{\nu}_4 \gamma^\mu \nu_4,
\end{equation}
where the Dirac dark matter field is given by $\nu_4=\tilde{\nu}_{4L}+\nu_{4R}$ with a Dirac mass 
$m_{\nu}\overbar{\nu}_4  \nu_4$ where we have defined $m_{\nu}\equiv M^{\nu}_4$.

\vspace{1cm}

We finish this section by summarizing all non-SM interactions that will later be relevant for our phenomenological analysis, introducing the notation that we shall subsequently use: 
\begin{equation}
{\cal L} \supset Z'_\mu \left(
g_{bb} \bar{q}_{L} \gamma^\mu q_{L} 
+ g_{bs}\bar{b}_{L} \gamma^\mu s_{L} 
+ g_{\mu \mu}\bar{\ell}_{L} \gamma^\mu \ell_{L} 
+ g_{\nu \nu} \overbar{\nu}_4 \gamma^\mu \nu_4
\right),
\label{eq:Zp_Rk_couplings}
\end{equation}
where $q_L = (t_L,b_L)^T$, $\ell_L = (\nu_{\mu \, L}, \mu_L)^T$, 
$g_{bb}=g'q_{Q4}(s^Q_{34})^2$, $g_{bs}=g_{bb} (V'^\dagger_{dL})_{32}$, $g_{\mu \mu}=g'q_{L_4}(s^L_{24})^2$, $g_{\nu \nu} =  g'q_{\nu_4}$.
We expect $|(V'^\dagger_{dL})_{32}| \lesssim |V_{ts}|$, where $|V_{ts}|\approx 0.04$ is the 3-2 entry of the CKM matrix, as otherwise unnatural cancellations would be required.  
It follows that $|g_{bs}| \lesssim  |V_{ts}g_{bb}|$;  
in the following for simplicity we assume  $g_{bs} = V_{ts} g_{bb}$, and that $g_{bb}$ and $g_{\mu \mu}$ have the same sign. 
Thus, the relevant parameter space is 5-dimensional: 3 couplings ($g_{bb}$, $g_{\mu \mu}$, $g_{\nu \nu}$) and 2 masses ($M_{Z'}$ and the dark matter mass $m_{\nu}$).  
From the theory point of view these are all essentially free parameters, 
although one naturally expects $g_{\nu \nu} \gg  g_{bb}, g_{\mu \mu}$ in the absence of large mixings or large  hierarchies of $U(1)'$ charges.   
These parameters are then constrained by flavour physics, multiple low-energy precision measurements, colliders, and dark matter detection experiments.
In the following sections we work out these constraints, and identify the regions of the parameter space where both the B-anomalies and the dark matter relic abundance can be explained without conflicting any existing experimental data.  
We note that $Z^{\prime}$ models simultaneously addressing the B-anomalies and dark matter have been previously discussed in Refs.~\cite{Sierra:2015fma,Belanger:2015nma,Celis:2016ayl,Altmannshofer:2016jzy,Baek:2017sew,Cline:2017qqu,Fuyuto:2017sys}.   
In particular, Ref.~\cite{Altmannshofer:2016jzy} performed a detailed analysis of collider, precision, dark matter constraints in a similar model based on gauged $L_\mu-L_\tau$ symmetry. 
The main practical difference between our setup and that model is the presence of $Z^{\prime}$ couplings to b-quarks in Eq.~\ref{eq:Zp_Rk_couplings}, which affects the LHC phenomenology as well as direct and indirect detection signals.  

%%%%%%%%%%%%%%%%%%%%%%%%%%%%%%%%%%%%%%%%%%%%
\section{$\mathbf{R_{K^{(*)}}}$ anomalies and flavour constraints}
\label{constraints}

In this section we review and update the constraints on the parameter space of $Z'$ models motivated by the current B-meson anomalies.      
One possible explanation of the $R_K$ and $R_{K^*}$ measurements in LHCb is that the low-energy Lagrangian below the weak scale contains an additional contribution to the effective 4-fermion operator with left-handed muon, $b$-quark, and $s$-quark fields:  
\begin{equation}
\Delta \mathcal{L}_\text{eff} \supset G_{b s\mu} ( \bar{b}_L \gamma^\mu s_L) (\bar{\mu}_L \gamma_\mu \mu_L )+ {\rm h.c.}, \qquad G_{b s\mu} \approx \frac{1}{(31.5\text{ TeV})^2}. 
\label{eq:bsmu_fit}
\end{equation}
Above, the numerical value of the effective coefficient corresponds to the best fit quoted in Ref.~\cite{Altmannshofer:2017yso}. 
In our model, this operator arises from tree-level $Z'$ exchange  and the analogous operator with $\mu_L$ replaced by $e_L$ does not appear due to vanishing charged lepton mixing.  
We can express the coefficient $G_{b s\mu}$ as function of the couplings in Eq.~\ref{eq:Zp_Rk_couplings}, 
\begin{equation}
G_{b s\mu}= - \frac{g_{bs}g_{\mu\mu}}{M^2_{Z'} }= - \frac{V_{ts} g_{bb}g_{\mu\mu}}{M^2_{Z'}}. 
\label{eq:bsmu_model}
\end{equation}
Together, Eqs.~(\ref{eq:bsmu_fit}) and (\ref{eq:bsmu_model}) imply the constraint on the parameters 
$g_{bb}$, $g_{\mu\mu}$ and $M_{Z'}$:  
\begin{equation}
\frac{g_{bb}g_{\mu\mu}}{M^2_{Z'}} \approx \frac{1}{(6.4\text{ TeV})^2}.
\label{eq:coeff_RK}
\end{equation} 

There are additional constraints on these parameters coming from flavour physics and low-energy precision measurements. 
In the following we  determine the region of the parameter space where the $R_{K^{(*)}}$ anomalies can be explained without conflicting other experimental data.

\noindent{\bf \underline{$B_s-\overbar{B}_s$ mixing}}

\begin{figure}
\begin{center}
\includegraphics[width=12cm]{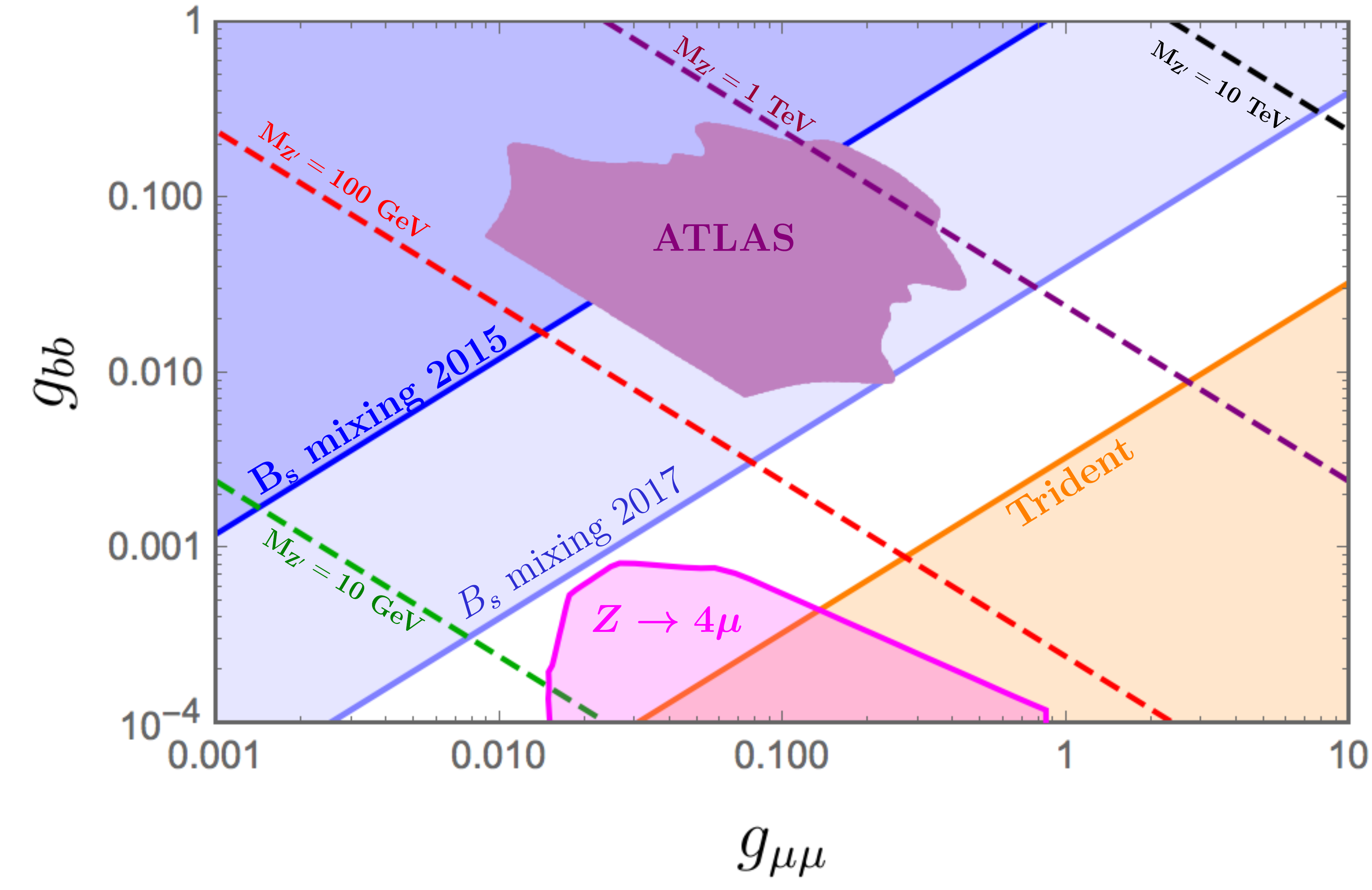}
\caption{
The parameter space in the $(g_{\mu\mu},g_{bb})$ plane compatible with \RK anomalies and flavour constraints (white). 
The $Z'$ mass varies over the plane, with a unique $Z'$ mass for each point in the plane as determined by Eq.~\ref{eq:coeff_RK}. 
We show the recent $B_s$ mixing constraints (light blue), and the trident bounds (orange);   
for reference we also display the previous weaker $B_s$ mixing bounds (dark blue). 
The green, red, purple and black lines correspond to $M_{Z'}=10, 100, 1000, 10000 \text{ GeV}$ respectively.
} 
\label{fig:constrained_couplings}
\end{center}
\end{figure}

The $Z^{\prime}$ coupling to $bs$ leads to an additional tree-level contribution to $B_s-\overbar{B}_s$ mixing. Low-energy observables are affected by the effective operator arising from integrating out the $Z'$ at tree level: 
\begin{equation}
\Delta \mathcal{L}_\text{eff} \supset - {G_{b s}\over 2}  (\bar{s}_L \gamma^\mu b_L)^2 + {\rm h.c}, 
\qquad G_{b s} = \frac{g_{bs}^2}{M^2_{Z'}} = \frac{g_{bb}^2 V^2_{ts}}{M_{Z'}^2}.
\label{eq:effective_bs}
\end{equation}
Such a new contribution is highly constrained by the measurements of the mass difference $\Delta M_s$ of neutral $B_s$ mesons.
In this paper we follow the recent analysis of Ref.~\cite{DiLuzio:2017fdq} which, using updated lattice results, obtains a stronger bound on $G_{b s}$: 
\begin{equation}
- \frac{1}{(180\text{ TeV})^2}  \lesssim G_{b s} \lesssim \frac{1}{(770\text{ TeV})^2}, 
\qquad @ \, 95\% {\rm CL}.
\end{equation}
The resulting constraints in the $(g_{\mu\mu},g_{bb})$ plane are shown as the light blue region in  Fig.~\ref{fig:constrained_couplings}. 
The updated constraint is particularly strong for the models that generate a strictly positive $G_{b s}$ \cite{DiLuzio:2017fdq} (as is the case in $Z'$ models) due to the $\sim 1.8\sigma$ discrepancy between the measured $\Delta M_s$ and the updated SM predictions which favors $G_{b s} < 0$.   
As a consequence, $Z'$ models explaining the B-meson anomalies required $M_{Z'} \lesssim 1$~TeV, 
assuming weak coupling $g_{\mu\mu} \lesssim 1$. 
For easy reference, we also show the $B_s$ mixing constraints based on the previous SM determination of $\Delta M_s$ \cite{Artuso:2015swg},
$- \frac{1}{(160\text{ TeV})^2}  \lesssim G_{b s} \lesssim \frac{1}{(140 \text{ TeV})^2}$, 
see the dark blue region in Fig.~\ref{fig:constrained_couplings} labeled ``$B_s$ mixing 2015''.

\noindent{\bf \underline{Neutrino trident}}

The $Z'$ coupling to left-handed muons leads to a new tree-level contribution to the effective 4-lepton interaction  
\begin{equation}
\Delta \mathcal{L}_\text{eff} \supset - {G_{\mu}\over 2}  (\bar \ell_L \gamma^\mu \ell_L)^2, 
\qquad G_{\mu} = \frac{g_{\mu\mu}^2}{M^2_{Z'}}.
\label{eq:effective_mu}
\end{equation}
This operator is constrained by the trident production $\nu_\mu \gamma^* \rightarrow \nu_\mu \mu^+ \mu^-$ \cite{Geiregat:1990gz,Mishra:1991bv,Altmannshofer:2014pba}. 
Using the results of the global fit in Ref.~\cite{Falkowski:2017pss}, the bound on the effective coefficient is given by    
\begin{equation}
-  \frac{1}{(390\text{ GeV})^2} \lesssim G_{\mu} \lesssim \frac{1}{(370\text{ GeV})^2}, 
\qquad @ \, 95\% {\rm CL}.
\label{eq:effective_trident}
\end{equation}
The limits in the $(g_{\mu\mu},g_{bb})$ plane are shown as the orange region in  Fig.~\ref{fig:constrained_couplings}.
Since the trident constraints probe much lower scales than the $B_s$ mixing, a much larger $Z'$ coupling to muons is allowed, $g_{\mu \mu} \gtrsim 1$ for a heavy enough $Z'$. 
Nevertheless, together with the $B_s$ mixing constraints, the trident leaves only a narrow sliver of the parameter space that could address the $B$ meson anomalies.

\noindent{\bf \underline{LHC searches}}

Further constraints on our model come from collider searches. 
For light $Z'$ masses, the LHC measurements of the Z decays to four muons, with the second muon pair produced in the SM via a virtual photon~\citep{CMS:2012bw,Aad:2014wra}, $pp \rightarrow Z \rightarrow 4 \mu$, sets relevant constraints in the low mass region of $Z'$ models, $5\lesssim M_{Z'} \lesssim 70$ GeV. 
The $Z \rightarrow 4 \mu$  constraints on the magnitude of the  $Z'$ coupling to muons were analyzed in Refs.~\cite{Altmannshofer:2014cfa,Altmannshofer:2014pba,Altmannshofer:2016jzy}. 
Projecting these results onto our model, the excluded parameter space is marked as the pink regions in Fig.~\ref{fig:constrained_couplings} and in the upper-left panel of Fig.~\ref{fig:ATLASlimits}. 
All in all, the $Z \rightarrow 4 \mu$ constraint is non-trivial but for any $Z'$ mass it always leaves some available parameter space to explain the B-meson anomalies. 

For a heavier $Z'$, the strongest constraints comes from LHC dimuon resonance searches, $pp\rightarrow Z'\rightarrow \mu^+ \mu^-$, see also  \cite{Dalchenko:2017shg}. 
In our model the $Z'$ is dominantly produced at the LHC through its couplings to bottom quarks, $b \bar b \to Z'$. 
The cross section $\sigma (p p \to Z')$ from $b \bar b$ collisions is taken from Fig.~3 of Ref.~\cite{Faroughy:2016osc}.
The contribution of bottom-strange collisions, which is subleading in our model, is estimated using {\tt Madgraph}~\cite{Alwall:2014hca}. 
The $Z'$ boson can subsequently decay into muons, muon neutrinos, bottom or strange quarks, and also into top quarks and dark matter when kinematically allowed. 
The partial decay widths are given by 
\begin{equation}
\begin{split}
& \Gamma_{Z'\rightarrow \mu \bar{\mu}}=\frac{1}{24\pi}g_{\mu \mu}^2 M_{Z'}  
=  \Gamma_{Z'\rightarrow \nu_\mu \bar{\nu}_\mu}, \\
& \Gamma_{Z'\rightarrow b \bar{b}}=\frac{1}{8\pi}g_{bb}^2 M_{Z'},
\qquad \Gamma_{Z'\rightarrow b \bar{s}}=\frac{1}{8\pi}g_{bb}^2V_{ts}^2 M_{Z'},  \\
& \Gamma_{Z'\rightarrow t \bar{t}}=\frac{1}{8\pi}g_{bb}^2 M_{Z'} \left( 1-\frac{m_t^2}{M_{Z'}^2} \right) \sqrt{1-\frac{4 m_t^2}{M_{Z'}^2}}, \\
& \Gamma_{Z'\rightarrow \nu_4 \bar{\nu}_4}=\frac{1}{24\pi}g_{\nu\nu}^2 M_{Z'} \left( 1-\frac{m_\nu^2}{M_{Z'}^2} \right) \sqrt{1-\frac{4 m_\nu^2}{M_{Z'}^2}}, \\
\end{split}
\end{equation}
from which we  calculate ${\rm Br}(Z' \to \mu \mu)$ analytically. 
Then $\sigma (p p \to Z' \to \mu \mu)$ is estimated using the narrow-width approximation, and compared with the limits from the recent dimuon resonance search  by ATLAS~\cite{Aaboud:2017buh}, which allows us to constrain $Z'$ mases between $150$ GeV and $5$ TeV. 
We verified that the analogous  Tevatron analyses give weaker constraints, also in the low mass regime. 
Fig.~\ref{fig:ATLASlimits} shows the ATLAS constraints for specific $Z'$ masses (200, 500 and 1000 GeV) with dark matter couplings set to zero and arbitrary $(g_{\mu \mu}, g_{bb})$ couplings.   
Fig.~\ref{fig:constrained_couplings} shows the same limits for the $Z'$ mass fixed in function of  $(g_{\mu \mu}, g_{bb})$ by the condition in Eq.~\ref{eq:coeff_RK}. 
We conclude that in the  parameter space of our model relevant for explaining the B-meson anomalies   
the ATLAS dimuon limits are always weaker that the new $B_s$ mixing constraints. 

\noindent{\bf \underline{Constraints from lepton-flavour violation}}
% $\mu \rightarrow e \gamma$

\begin{figure}
\begin{center}
\includegraphics[width=12cm]{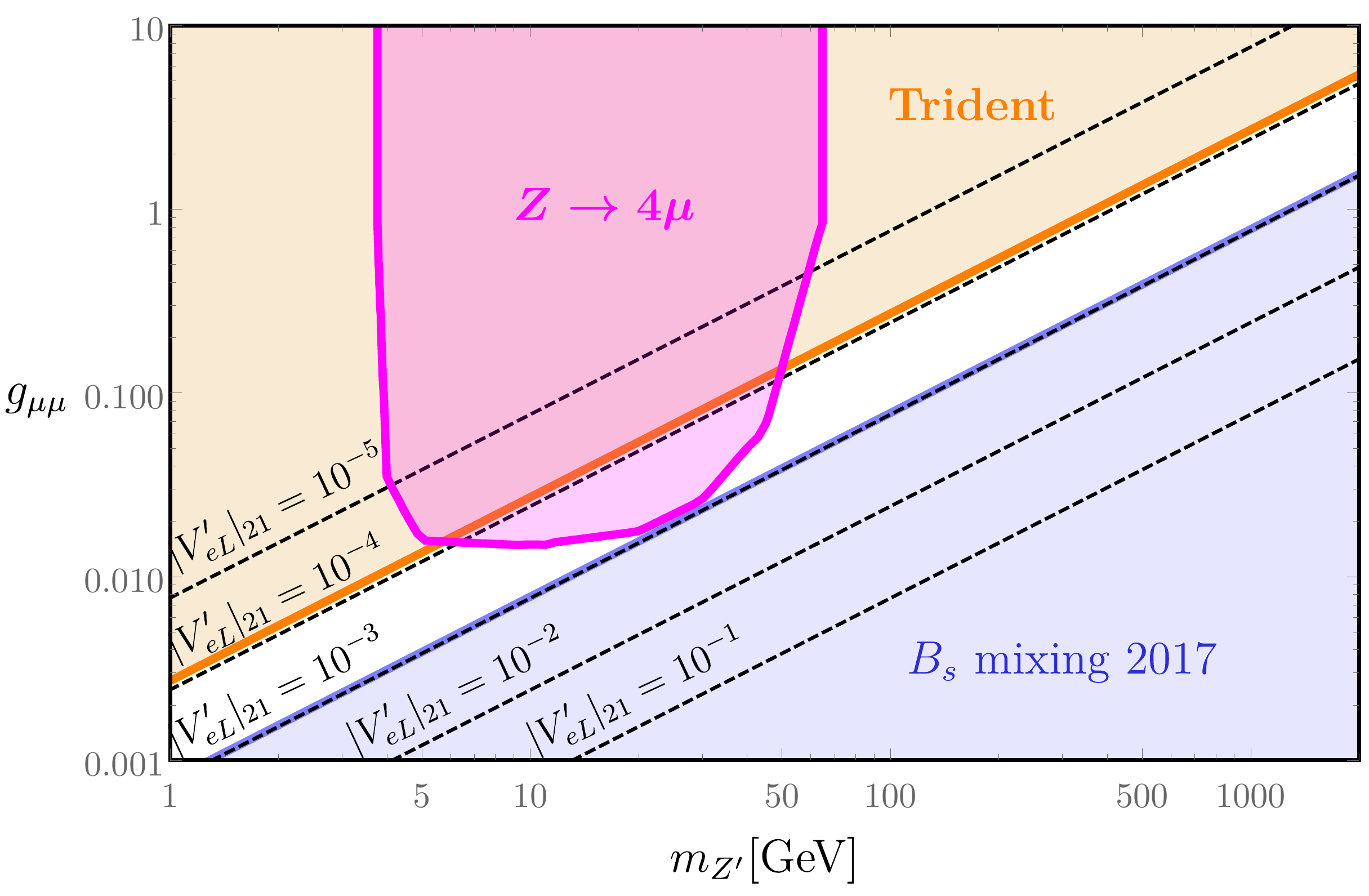}
\caption{
The parameter space in the $(g_{\mu\mu},m_{Z^\prime})$ plane compatible with \RK anomalies and flavour constraints (white). 
We show the recent $B_s$ mixing constraints (light blue), the trident bounds (orange), the $Z\rightarrow 4\mu$ limits (pink) as well as the expected limits from $\mu \rightarrow e \gamma$ for several values of $|V'_{eL}|_{21}$ (black dashed).
} 
\label{fig:mutoegamma}
\end{center}
\end{figure}

\begin{figure}
\begin{minipage}[b]{0.5\textwidth}
\includegraphics[width=7.6cm]{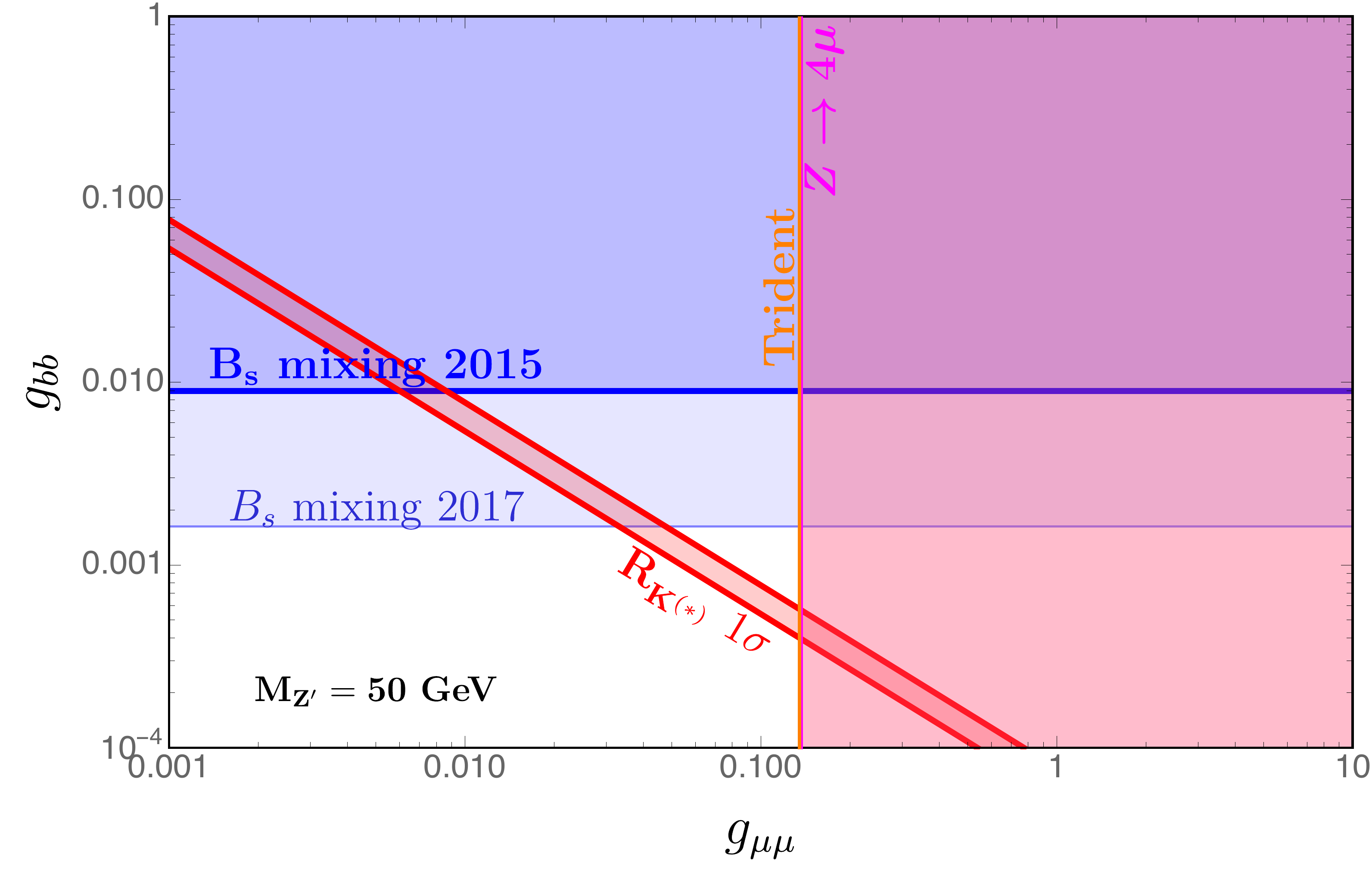} 
\end{minipage}
\begin{minipage}[b]{0.5\textwidth}
\includegraphics[width=7.6cm]{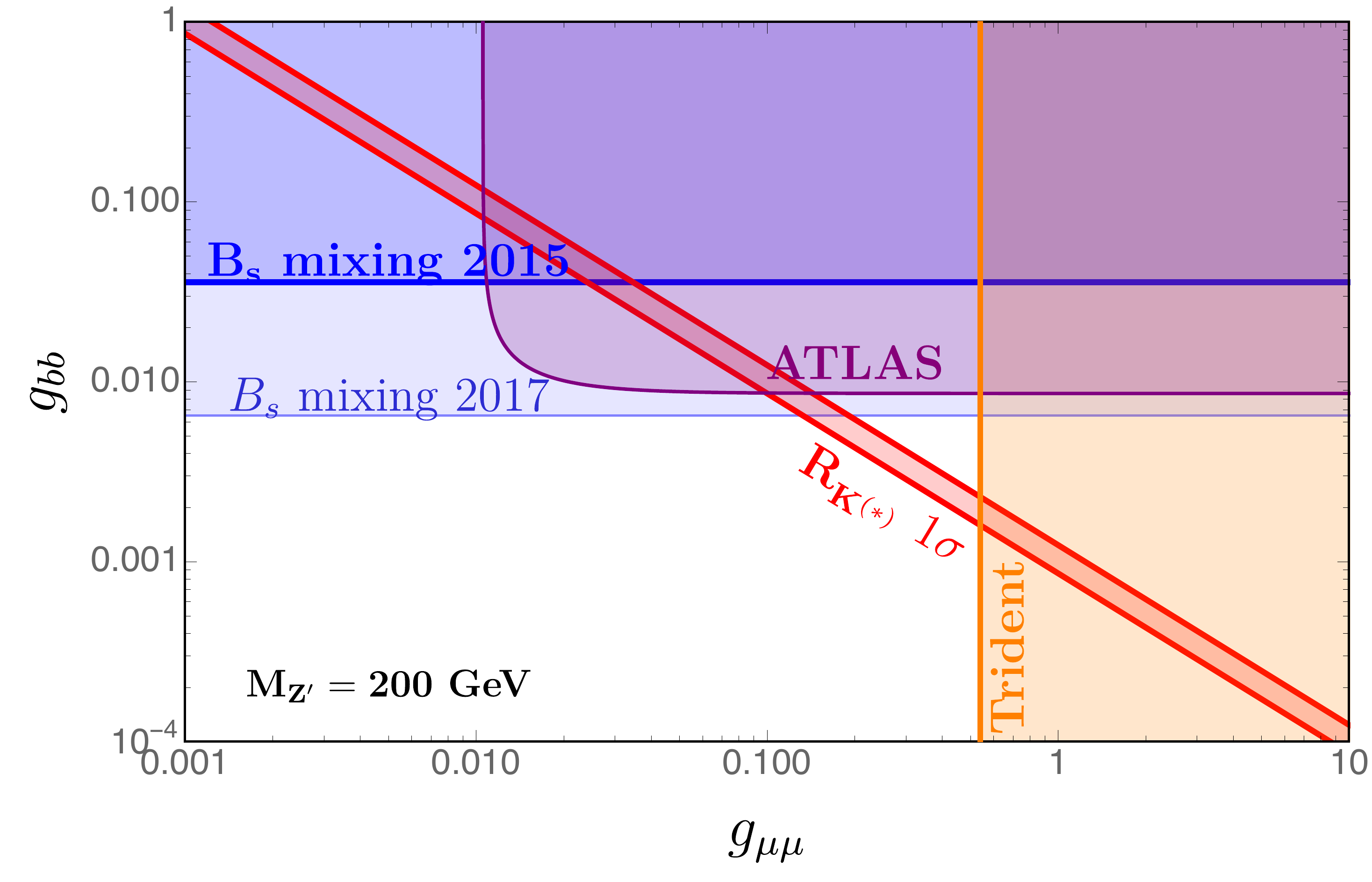}
\end{minipage}
\begin{minipage}[b]{0.5\textwidth}
\includegraphics[width=7.6cm]{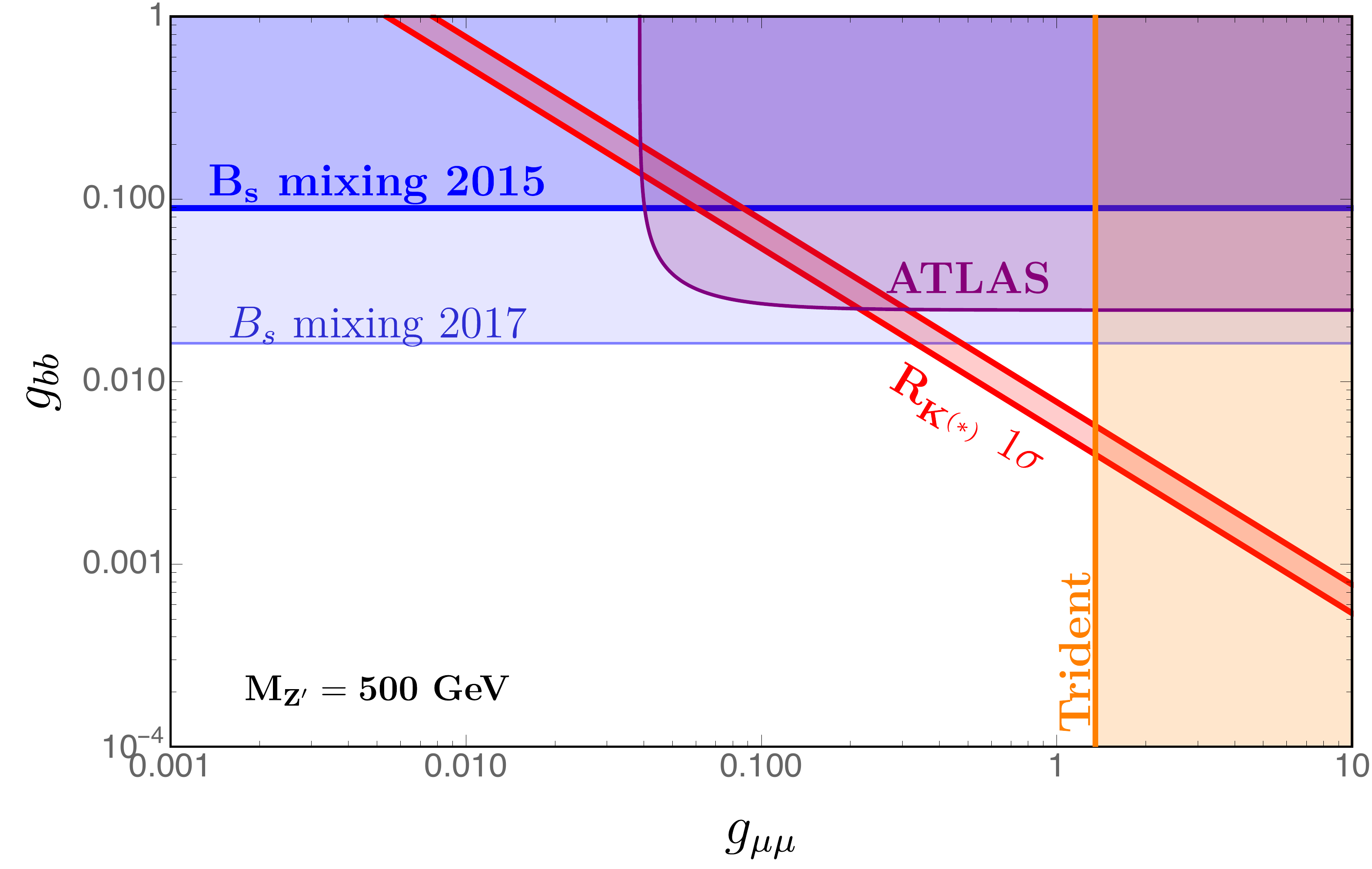}
\end{minipage}
\begin{minipage}[b]{0.5\textwidth}
\includegraphics[width=7.6cm]{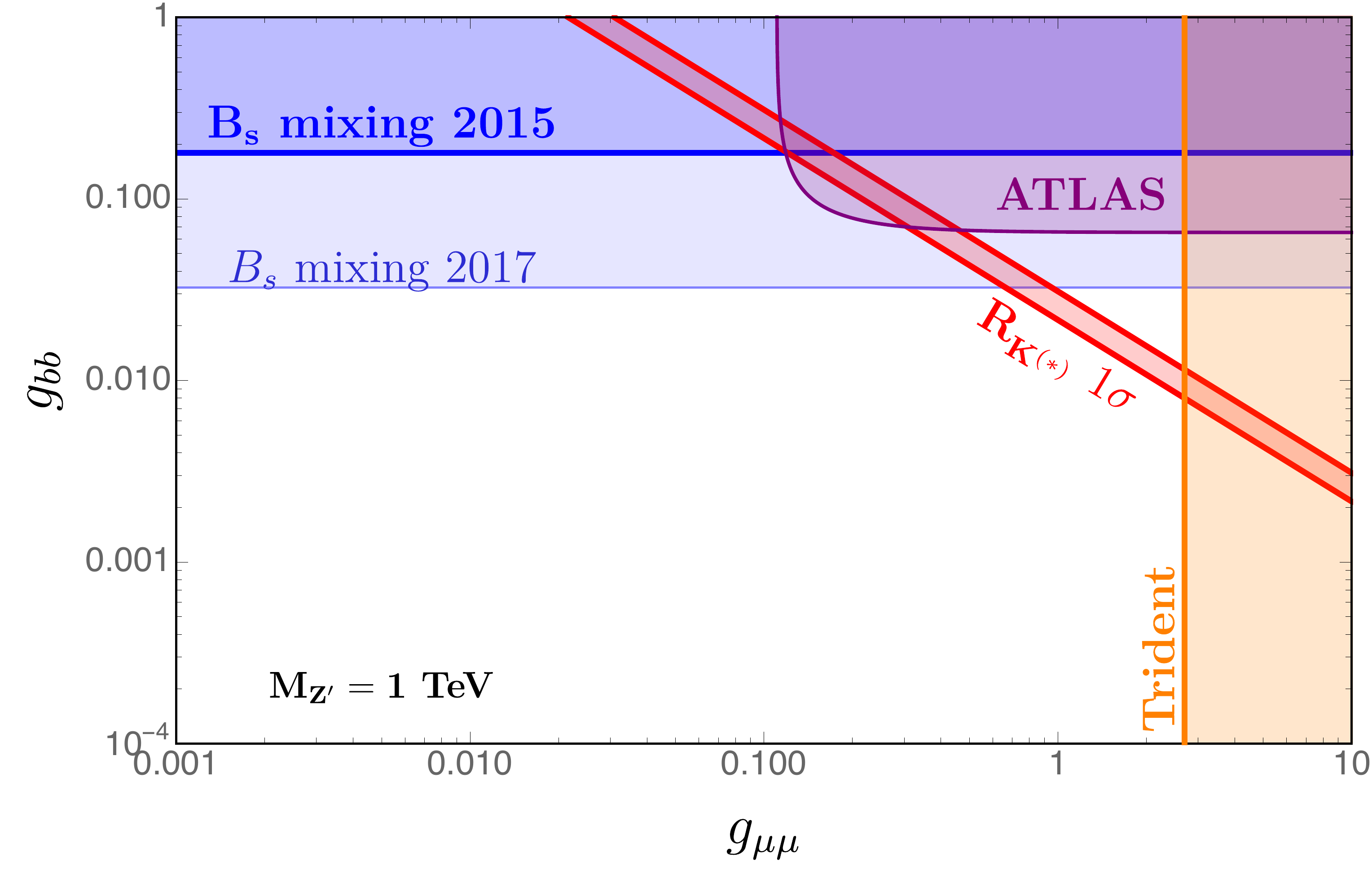}
\end{minipage}
\caption{ 
Bounds on the parameter space in the $(g_{\mu\mu},g_{bb})$ plane for fixed $Z'$ masses: 50, 200, 500 and 1000 GeV, as indicated on each panel. The red bands explain \RK at $1\sigma$. The blue and orange areas show the $B_s-\overbar{B}_s$ mixing~\cite{DiLuzio:2017fdq} and neutrino trident~\cite{Falkowski:2017pss} $2\sigma$ exclusions, respectively. For low $Z'$ masses we have additional constraints from $Z\rightarrow4\mu$ as shown in pink. The ATLAS limits~\cite{Aaboud:2017buh} from dimuon resonance searches for $36 \text{ fb}^{-1}$luminosity  are given in purple for larger $Z'$ masses.}
\label{fig:ATLASlimits}
\end{figure}

So far we were assuming zero mixing in the charged-lepton sector. It is interesting to discuss the constraints resulting from relaxing that assumption. In particular, for a non-vanishing mixing angle between charged leptons of the second and first generations $(V'_{eL})_{21} \neq 0$, a non-diagonal $Z^\prime$ coupling to left-handed muons and electrons would be present
\begin{equation}
\mathcal{L}\supset g_{\mu \mu} (V'_{eL})_{21} \bar{\mu}_L \gamma^\mu e_L Z^\prime_\mu+\text{h.c.}~,
\end{equation}
which could generate an additional contribution to the transition $\mu \rightarrow e \gamma$ whose partial decay width can be estimated, according to~\cite{Raby:2017igl}, as 
\begin{equation}
\Gamma (\mu \rightarrow e \gamma)\simeq \dfrac{\alpha ~ m_\mu^5}{1024 \pi^4 m_{Z^\prime}^4} g_{\mu \mu}^4 |V'_{eL}|_{21}^2 F^2(m_\mu^2/m_{Z^\prime}^2)~,
\end{equation}
where $F(x)$ is a loop function, as defined in~\cite{Raby:2017igl}, whose limit for $m_{Z^\prime} \gg m_\mu$ is $\lim_{x \to 0} F(x) = 2/3$. The branching ratio of $\mu \rightarrow e \gamma$ is severely constrained by the MEG experiment~\cite{TheMEG:2016wtm} which set the bound $\text{BR}(\mu \rightarrow e \gamma)\leq
 4.2\times 10^{-13}$ at $90\%$CL. An analytical approximation of this branching ratio is given by
 \begin{equation}
\text{BR}(\mu \rightarrow e \gamma) \simeq 1.24 \times 10^{-6} ~g_{\mu \mu}^4 |V'_{eL}|_{21}^2  \left(\dfrac{m_{Z^\prime}}{1~\text{TeV}} \right)^{-4}~,
 \end{equation}
implying that $\mu \rightarrow e \gamma$ is expected to set a stronger constraint than the neutrino trident production for values of the mixing angle $|V'_{eL}|_{21} \gtrsim 10^{-4} $ as represented in Fig.~\ref{fig:mutoegamma}, while $|V'_{eL}|_{21}\gtrsim 10^{-3}$ would rule out the entire parameter space. As a result, in the viable parameter space of our setup, the mixing angle $|V'_{eL}|_{21}$ is expected to be $|V'_{eL}|_{21}\lesssim 10^{-4}$.
Similarly, the experimental limit on the lepton-flavour-violating of the tau lepton into 3 muons, ${\rm Br}(\tau \to 3 \mu) \leq 2\times 10^{-8}$  \cite{Hayasaka:2010np}, constrains the mixing angle between charged leptons of the second and third generation $(V'_{eL})_{32}$: 
${g_{\mu \mu}^2 |V'_{eL}|_{32} \over m_{Z^\prime}^2} \lesssim {1 \over (16~{\rm TeV})^2}$. 
This is stronger than the trident bound in Eq.~(\ref{eq:effective_trident}) for $(V'_{eL})_{32} \gtrsim 3 \times 10^{-4}$, while $(V'_{eL})_{32} \gtrsim 3 \times 10^{-3}$ would rule out the entire parameter space.

\noindent{\bf \underline{Other constraints}}

Finally we comment on other precision observables which yield subleading constraints on our model.

The  contribution of $Z'$ to the muon magnetic moment is given by 
\begin{equation}
\Delta^\mu_{g-2} = {1 \over 12 \pi^2} m_\mu^2 \left (g_{\mu \mu} \over M_{Z'} \right )^2 . 
\end{equation}
The measured discrepancy of the muon magnetic moment is 
$\Delta^\mu_{g-2} = (290 \pm 90)\times 10^{-11}$ \cite{Jegerlehner:2009ry}. 
This sets weaker limits on the ratio $g_{\mu \mu}/M_{Z'}$ than the trident production. 

Next, $Z'$ exchange generates the effective interaction between b-quarks and muons: 
\begin{equation}
\mathcal{L}_\text{eff} \supset  G_{b\mu} (\bar{b}_L \gamma^\mu b_L)(\bar{\mu}_L\gamma_\mu \mu_L),
 \qquad G_{b\mu} = - \frac{g_{bb} g_{\mu \mu}}{M^2_{Z'}} =  -  \frac{1}{(6.4\text{ TeV})^2},
\label{eq:op_upsilon}
\end{equation}
where we used Eq.~\ref{eq:coeff_RK}. 
The operator in Eq.~\ref{eq:op_upsilon} is constrained by lepton flavour universality of upsilon meson decays \cite{Aloni:2017eny}. 
Focusing on the $\Upsilon_{1s}$ state, given the measured ratio~\cite{Patrignani:2016xqp} 
\begin{equation}
R^{\tau/\mu}_{1s}=\frac{\Gamma(\Upsilon_{1s}\rightarrow \tau^+ \tau^-)}{\Gamma(\Upsilon_{1s}\rightarrow \mu^+\mu^-)}=1.008 \pm 0.023,
\end{equation}
and the SM prediction is $R^{\tau/\mu}_{1s}=0.9924$, one finds the constraint 
\begin{equation}
-\frac{1}{(150\text{ GeV})^2} < G_{b\mu} < \frac{1}{(190\text{ GeV})^2}
\qquad @ \, 95\% {\rm CL}.
\end{equation}
This is automatically satisfied in our model in the parameter space where the \RK anomalies are explained.

\section{Dark Matter}
\label{darkmatter}

Our model  comprises a fourth neutrino ($\nu_4$) which  possesses all the properties of a viable dark matter candidate. Indeed, $\nu_4$ is an electrically neutral particle interacting weakly with the SM sector through an exchange of $Z'$. 
Furthermore, our charge assignments under the local symmetries forbid any mixing with other fields such as the SM neutrinos. 
Therefore, conservation of fermion number in the dark sector can be effectively seen as a $\mathbb{Z}_2$ symmetry forbidding the dark matter from decaying and as a consequence ensuring its stability. 
In this section we discuss in some detail the generation of the dark matter relic density and show the constraints  from indirect and direct dark matter searches.
For brevity, in the following the dark matter candidate is simply denoted as $\nu$.

\subsection{Relic abundance} 
\begin{figure}[t]
\begin{center}
\includegraphics[width=12 cm]{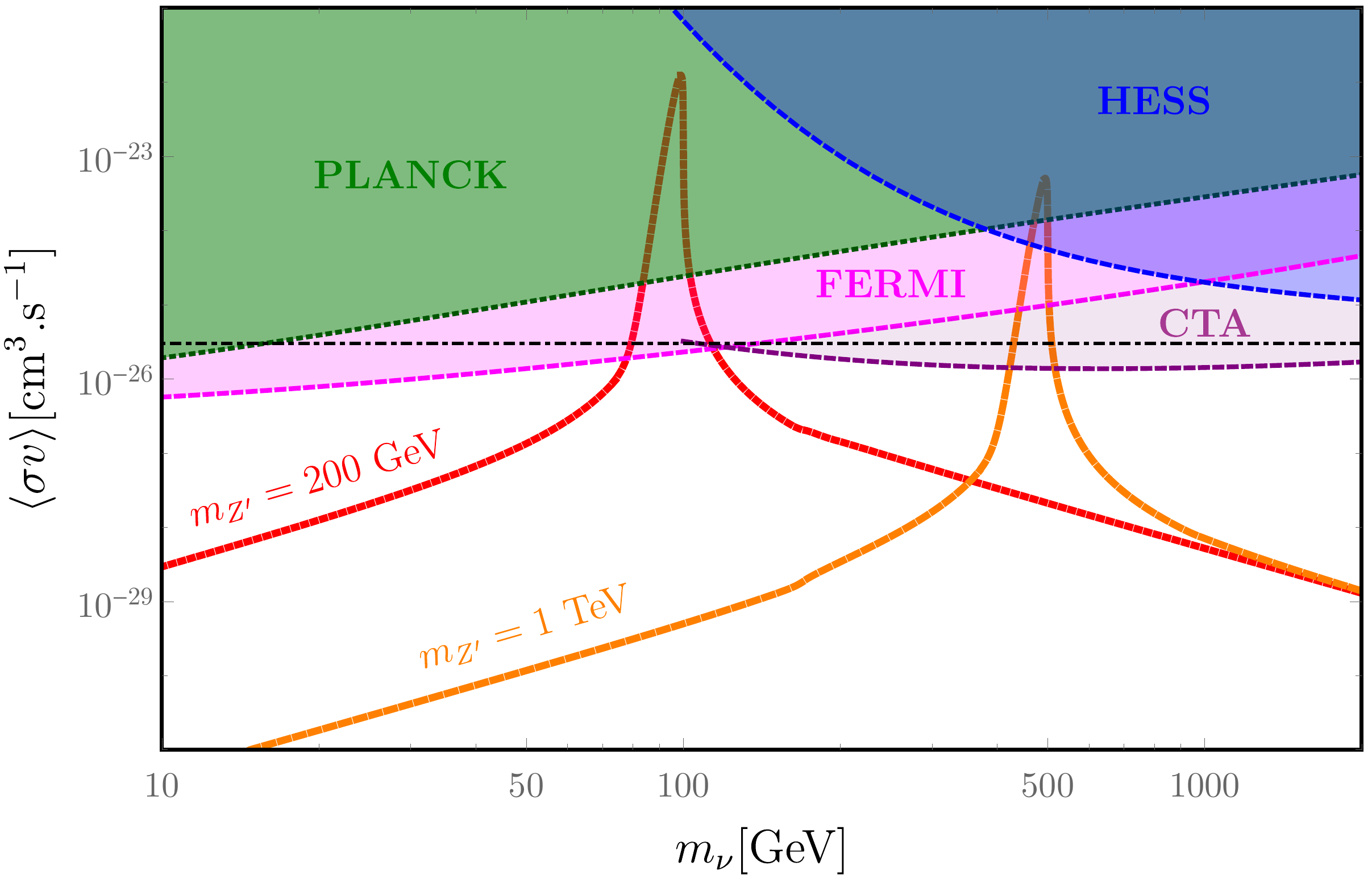}
\caption{Dark matter velocity averaged annihilation cross section for $M_{Z^\prime}=200(1000)$ GeV in red (orange) assuming $g_{bb}=g_{\nu \nu}=g_{\mu \mu}=0.1$ and indirect detection limits assuming $\bar{b}b$ as final state from HESS~\cite{Abdallah:2016ygi} in blue, Fermi~\cite{Ackermann:2015zua} in pink and predictions for the upcoming CTA~\cite{Wood:2013taa} assuming 500h of observation toward the Galactic Center in purple. Limits from the Planck collaboration~\cite{Slatyer:2015jla} are shown in green, assuming dark matter annihilation to $\bar{\mu} \mu$. The dotted-dashed black line represents the canonical value of the cross section $\langle \sigma v \rangle =3 \times 10^{-26} \text{cm}^3~\text{s}^{-1}$.} 
\label{fig:relic_density_ID}
\end{center}
\end{figure}

Our dark matter candidate is a weakly interacting massive particle (WIMP) whose relic abundance can be generated via the well studied freeze-out scenario~\cite{Bertone:2004pz,Arcadi:2017kky}. 
We will fit the parameters to reproduce the present dark matter density measured by the Planck collaboration: $\Omega_\text{DM} h^2 = 0.1198 \pm{0.0015}$~\cite{Ade:2015xua}. 
In the WIMP scenario, $\Omega_\text{DM}$ is determined by the thermally averaged annihilation cross section $\langle \sigma v \rangle$ which can be expressed as~\cite{Gondolo:1990dk}
\begin{equation}
\left< \sigma v \right> 
=\frac{1}{4 x^4 K_2(x)^2}\int^\infty_{2x} \text{d} z (z^2-4x^2)z^2 K_1(z) 
\sigma_{s = z^2 T^2} ~,
\label{sigmavgondolo}
\end{equation}
where $x =  m_{\nu}/T$, %  $z =  \sqrt{s}/T$, 
$K_{1,2}(x)$ are the modified Bessel functions of the second kind, 
 and $\sigma_s$ is the dark matter annihilation cross section at the centre-of-mass energy squared $s$. 
When $\langle  \sigma v \rangle $ is approximately independent of the temperature the relic density is related to it as  
\begin{equation}
\Omega_\text{DM} h^2 \simeq 0.12
\dfrac{\langle \sigma v \rangle_{\rm thermal}}{\langle \sigma v \rangle}~, \qquad 
\langle \sigma v \rangle_{\rm thermal} \equiv 3 \times 10^{-26}\text{cm}^3~\text{s}^{-1}. 
\end{equation}
In our model the dark matter particles can annihilate to\footnote{%
Since $g_{bs} \ll g_{bb}$, we can safely ignore annihilation to $\bar{b}s$ and $\bar{s}b$.
}
$\bar{\mu}\mu, \bar{\nu}_\mu \nu_\mu, \bar{b}b$, and possibly to $\bar{t}t,Z^\prime Z^\prime$, if kinematically accessible:   
\begin{equation}
\langle \sigma v \rangle = \sum_{\psi=b,t,\mu,\nu_\mu} \langle \sigma v \rangle_{\bar{\nu} \nu \rightarrow \bar{\psi} \psi} + \langle \sigma v \rangle_{\bar{\nu} \nu \rightarrow Z^\prime Z^\prime} . 
\end{equation}
One can derive an analytical approximation of $\langle \sigma v \rangle$ by expanding it in powers of $x^{-1}$ around the typical freeze-out temperature $x_\text{F} \sim 23$.
Away from the pole and thresholds, each component of $\langle \sigma v \rangle$ can be approximated by the s-wave expression:   
\begin{align}
\label{eq:sigmavapp}
\langle \sigma v \rangle_{\bar{\nu} \nu \rightarrow \bar{\psi} \psi}  \simeq &
\left \{ \begin{array}{ccr}
c_\psi \frac{g_{\nu \nu}^2 g_{\psi \psi}^2 }{4\pi} \frac{  m_{\nu}^2 }{M_{Z'}^4} & \hspace{2cm} &[M_{Z^\prime} \gg m_{\nu} \gg m_\psi ] ~
\\
c_\psi \frac{g_{\nu \nu}^2 g_{\psi \psi}^2 }{64\pi m_{\nu }^2}  & \hspace{2cm}   &[ m_{\nu} \gg M_{Z^\prime} \gg m_\psi ] ~ 
\end{array} \right . ,
\nonumber \\
\langle \sigma v \rangle_{\bar{\nu} \nu \rightarrow Z^\prime Z^\prime}  \simeq & \frac{ g_{\nu \nu}^4}{32 \pi  m_{\nu}^2} \hspace{2cm}   [ m_{\nu} \gg M_{Z^\prime} ] ~,
\end{align}
where $c_\psi$ is a color factor. 
One can see that the annihilation cross section grows as $m_{\nu}^2$ for small dark matter masses, and evolves as $m_{\nu}^{-2}$ for large dark matter masses.  
Therefore, for fixed couplings and $M_{Z'}$, there are typically two possible values of $m_{\nu}$ reproducing $\langle \sigma v \rangle_{\rm thermal}$, as illustrated in Fig.~\ref{fig:relic_density_ID}. 
For small couplings, $g \lesssim 0.1$, the annihilation cross section is substantially lower than the thermal one  except in the pole region, 
and the two solutions approach $m_{\nu} \sim M_{Z^\prime}/2$. 
As demonstrated in~\cite{Griest:1990kh}, the presence of a pole in the annihilation cross section may invalidate the $1/x$ expansion. 
In such a case one cannot use  Eqs.~\ref{eq:sigmavapp} and instead one has to rely on numerical evaluations using Eq.~\ref{sigmavgondolo}.
In order to explore the complete available parameter space, we compute the relic density and $\langle \sigma v \rangle$ numerically using the package micrOMEGAs~\cite{Belanger:2013oya} after implementing the model in FeynRules~\cite{Alloul:2013bka}.
For higher values of the couplings, $g \gtrsim 1$, the correct relic density can be achieved away from the pole region where Eqs.~\ref{eq:sigmavapp} are adequate.

\subsection{Indirect detection constraints}
In the WIMP framework, dark matter annihilations to SM states occurring inside large astrophysical structures such as the galactic center, dwarf spheroidal (dSphs) galaxies or galaxy clusters might be relatively frequent at the present time. This could lead to indirect dark matter observation by detecting the by-products of these annihilations in high-energy cosmic rays~\cite{Gaskins:2016cha}. 
The absence of any significant signal so far allowed experimental collaborations to derive upper limits on $\langle \sigma v \rangle$ as a function of the dark matter mass, assuming a particular dark matter density distribution for these astrophysical structures. 
In our model  $\langle \sigma v \rangle$ is approximately velocity independent in the non-relativistic limit, therefore the same value of order $\langle \sigma v \rangle_{\rm thermal}$ required to match the relic density is also relevant for indirect detection. 
The two annihilation channels most relevant for indirect detection are $\nu \bar \nu \to b \bar b$ and $\nu \bar \nu \to \mu^+ \mu^-$. 
In the parameter space where $\nu \bar \nu \to b \bar b$ (and possibly to $t \bar t$) dominates, the best current limits on $\langle \sigma v \rangle$ are derived by the Fermi-LAT collaboration from a combined analysis of 15 Milky Way dSphs and excludes dark matter masses $m_\text{DM} \lesssim 100$ GeV~\cite{Ackermann:2015zua}, 
assuming the Navarro-Frenk-White profile~\cite{Navarro:1995iw}. 
For larger dark matter masses stronger constraints on  the same annihilation channel come from the HESS experiment~\cite{Abdallah:2016ygi}, however the typical limits are  $\langle \sigma v \rangle \lesssim 10^{-25}\text{cm}^3~\text{s}^{-1}$ and therefore cross sections of order $\langle \sigma v \rangle_{\rm thermal}$ are not probed. 
In the future, sensitivity of the Cherenkov Telescope Array (CTA) might be sufficient to probe annihilation the thermal cross section for 
$m_\text{DM} \gtrsim 100$~GeV~\cite{Wood:2013taa,Pierre:2014tra,Silverwood:2014yza,Lefranc:2016dgx,Lefranc:2015pza}.
The current and future constraints in the $b \bar b$ annihilation channel are illustrated in Fig. \ref{fig:relic_density_ID}, where we also show predictions of our model for two particular points in the parameter space.   

As can be seen in Fig.~\ref{fig:constrained_couplings}, given the newer $B_s$ mixing constraints the allowed parameter space has $g_{\mu \mu} \gg g_{bb}$, and therefore annihilation into $\mu^+\mu^-$ (and the corresponding neutrinos) dominates. 
In such a case, the indirect detection limits on $\langle \sigma v \rangle$ are substantially weaker, such that the thermal annihilation cross section is allowed for dark matter masses above a few GeV \cite{Ackermann:2015zua}.  
For this reason, the indirect limits are not relevant in most of the interesting parameter space of our model. 
However, for small  dark matter masses $m_\nu \sim \text{GeV}$ annihilation  into leptons  at redshift $z \sim 1000$ can be constrained by CMB spectrum observations, as it  could modify the ionization history. 
For the thermal annihilation cross section, the Planck collaboration constraints on CMB spectrum distortions exclude dark matter masses below $m_\nu \lesssim 10 $ GeV~\cite{Slatyer:2015jla},  as illustrated in Fig.~\ref{fig:relic_density_ID}. 
We note that annihilation into leptonic final states can be relevant for experiments such as AMS-02  measuring cosmic-ray positrons and electrons, 
from which several studies have obtained strong constraints  on $\langle \sigma v \rangle$~\cite{Kopp:2013eka,Bergstrom:2013jra,Ibarra:2013zia,Lu:2015pta}. 
However these constraints are subject to strong dependence on the propagation model and uncertainties regarding cosmic-ray propagation in the interstellar medium, 
and for this reason we do not include them in the following.  
All in all, in our model dark matter masses $\lesssim 10$ GeV are excluded by the Planck collaboration results. Moreover, in the parameter space where annihilation into $b \bar b $ dominates, dark matter masses below 100 GeV are excluded by the Fermi-LAT results, although that parameter space is also disfavored by the recent $B_s$ mixing constraints.

\subsection{Direct Searches}

Direct detection (DD) of dark matter has proved to be extremely useful to constrain WIMP scenarios. These experiments aim at observing the recoil energy due to dark matter particles present in the Milky Way halo scattering off nuclei of a detector material. 
Their sensitivity has improved by several orders of magnitude during the past decade, 
and currently the xenon-based experiments LUX~\cite{Akerib:2016vxi}, PandaX~\cite{Tan:2016zwf} and XENON1T~\cite{Aprile:2017iyp} probe dark matter spin-independent (SI) scattering cross section of the order of $\sigma_\text{SI} \gtrsim 10^{-45}{\rm cm}^2$ for dark matter masses of the order of 100~GeV. 

\begin{figure}
\begin{minipage}[t!]{0.49\textwidth}
\includegraphics[height=5.3cm]{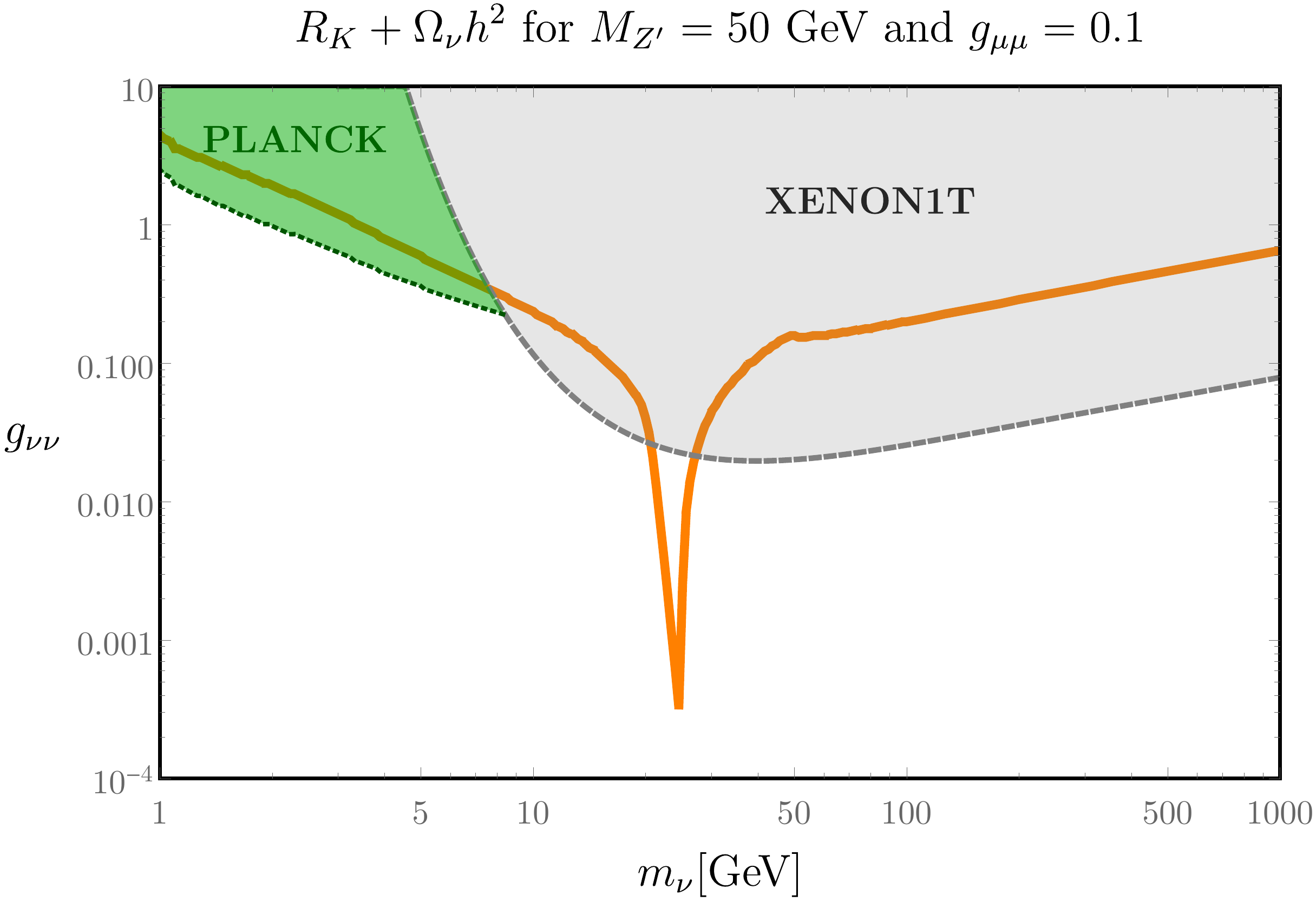}
\end{minipage}
\hfill
\begin{minipage}[t!]{0.49\textwidth}
\includegraphics[height=5.3cm]{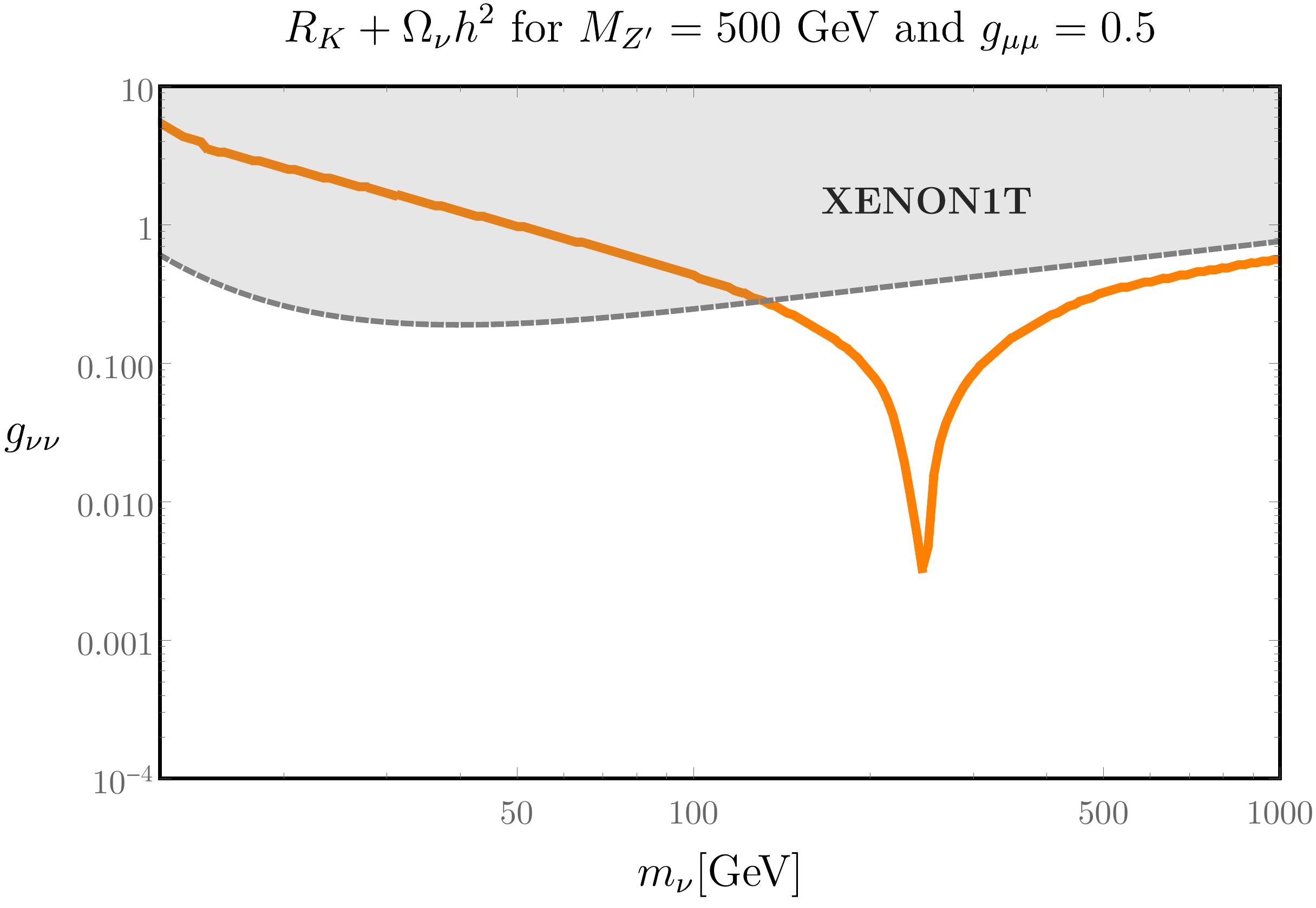}
\end{minipage}
\caption{Direct and indirect detection constraints on the parameter space : the orange line represents the model points featuring the correct dark matter  relic density and the appropriate Wilson coefficient explaining the $R_K$ discrepancy. The gray region shows the parameter space excluded by the XENON1T experiment~\cite{Aprile:2017iyp} and the green region represents the parameter space not excluded by direct detection experiments but in tension with the Planck collaboration~\cite{Slatyer:2015jla} results.}
\label{fig:DD}
\end{figure}

In our set-up, integrating out the tree-level $Z^\prime$ exchange between dark matter and the SM leads to the following effective operators at the scale $\mu \simeq M_{Z'}$:  
\begin{equation}
\mathcal{L}_\text{eff} \supset - \sum_{f=\mu,b} \dfrac{g_{\nu \nu} g_{f f}}{M_{Z^\prime}^2} \bar{f}_L \gamma^\alpha f_L  \bar{\nu} \gamma_\alpha \nu ~, 
\end{equation}
where again one can neglect the effective coupling to $b s$.  
Below the scale $M_{Z'}$, vector-like dark matter couplings to light quarks are induced via renormalization group (RG) running:  
\begin{equation}
\label{eq:Leffchiq}
\mathcal{L}_\text{eff} \supset \sum_{q = u,d} C_{1,f}^{(6)} (\mu) \bar{q} \gamma^\alpha q  \bar{\nu} \gamma_\alpha \nu ~ .  
\end{equation}
The complete RG equations can be found e.g. in \cite{DEramo:2014nmf}; schematically, one has $C_{1,f}^{(6)}(\mu) \sim {\alpha \over 4\pi} \frac{g_{\nu \nu} g_{ff}}{M_{Z^\prime}^2} \log\left (M_{Z\prime}  \over \mu \right )$.  
Other tensor structures beyond that in Eq.~\ref{eq:Leffchiq} also appear but they give subleading effects in direct detection.   
Finally, at $\mu \simeq 2$~GeV the couplings in Eq.~\ref{eq:Leffchiq} can be mapped to momentum- and spin-independent non-relativistic interactions of dark matter with protons and neutrons:  
\begin{equation}
\mathcal{L}_\text{eff,NR} \supset \sum_{N= p,n} c_1^N\bar{\nu} \nu \bar{N} N 
\end{equation}
where $c_1^p = 2 C_{1,u} + C_{1,d\chi}|_{\mu \simeq 2{\rm GeV}}$ and $c_1^n = C_{1,u} + 2 C_{1,d}|_{\mu \simeq 2{\rm GeV}}$. 
We evaluate numerically the one-loop RG evolution of effective couplings.  
To this end, above $m_Z$ we use the {\tt RunDM} package~\cite{DEramo:2016gos,DEramo:2014nmf,Crivellin:2014qxa}, while running below $m_Z$ and the coefficients $c_1^N$ are obtained by {\tt DirectDM}~\cite{Bishara:2017nnn}. 
For example, for $M_{Z'} = m_Z$ one finds  
\begin{equation}
c_1^p \simeq 3.1 \times 10^{-3}\left(\frac{g_{\mu \mu } g_{\nu \nu}}{M_{Z'}^2} \right) 
+ 2.5 \times 10^{-3}\left(\frac{ g_{bb} g_{\nu \nu}}{M_{Z'}^2}\right)~,
\qquad c_1^n = 0~, 
\end{equation}
while for $M_{Z'} = 1$~TeV: 
\begin{equation}
c_1^p \simeq 5.6 \times 10^{-3}\left(\frac{g_{\mu \mu } g_{\nu \nu}}{M_{Z'}^2} \right) 
+ 2.3 \times 10^{-3}\left(\frac{ g_{bb} g_{\nu \nu}}{M_{Z'}^2}\right)~,
\quad c_1^n \simeq 
4.5 \times 10^{-2}\left(\frac{ g_{bb} g_{\nu \nu}}{M_{Z'}^2}\right)~. 
\end{equation}
The coupling to neutrons vanishes within our approximations when $M_{Z'} \leq m_Z$.
For $M_{Z'} > m_Z$ a non-zero $c_1^n$ can be  generated, and is dominated by the top Yukawa contributions to  the RG running. 
The dark matter-nucleon spin-independent cross section can be straightforwardly derived from $\mathcal{L}_\text{eff,NR}$: 
\begin{equation}
\sigma^\text{N}_\text{DD} = \frac{ (c_1^N)^2 m_p^2 m_\nu^2}{\pi (m_p + m_\nu)^2}.   
\end{equation}
To compare with experimental bounds, which typically assume equal cross section on protons and neutrons, for a target nucleus with $Z$ protons and $A-Z$ neutrons we introduce the averaged cross section
\begin{equation}
\sigma_\text{DD} \simeq \frac{m_p^2 m_\nu^2}{\pi (m_p + m_\nu)^2}
{(Z c_1^p  + (A-Z)  c_1^n)^2 \over A^2}. 
\end{equation}
In the allowed parameter space relevant for the B-meson anomalies we have $g_{bb} \ll g_{\mu \mu}$. 
Assuming that  hierarchy, and also $m_\text{p} \ll m_\chi$, for xenon targets an approximate expression for the averaged cross section reads 
\begin{equation}
\sigma_\text{DD} \sim  
\Big( \dfrac{g_{\nu\nu}}{0.2} \Big)^2  
\Big( \dfrac{g_{\mu \mu}}{0.1} \Big)^2 
 \Big( \dfrac{m_Z}{M_{Z^\prime}} \Big)^{4} 
 10^{-45} ~\text{cm}^2 .
\end{equation}
In Fig.~\ref{fig:DD} we depicted the values of the $g_{\nu \nu}$ coupling satisfying the requirement of having the observed dark matter density as well at the correct value of the couplings $g_{\mu \mu}$ and $g_{bb}$ explaining the $R_K$ discrepancy. The left panel of that figure illustrates that, for low $M_{Z'}$, the XENON1T collaboration excludes dark matter masses away from the $Z^\prime$ pole but still allows for low dark matter masses $m_\nu \lesssim 10$~GeV. However, as discussed in the previous subsection, such low masses are excluded by the indirect Planck constraints, therefore the complementarity of direct and indirect detection searches indicates that the dark matter mass has to be close to the pole $m_\nu \sim M_{Z'}/2$. 
For larger $M_{Z'}$, dark matter masses away from the pole region are allowed, see the right panel of  Fig.~\ref{fig:DD}.

\section{Discussion and Conclusion}
\label{conclusion}

\begin{figure}
\begin{minipage}[t!]{0.49\textwidth}
\includegraphics[height=5.3cm]{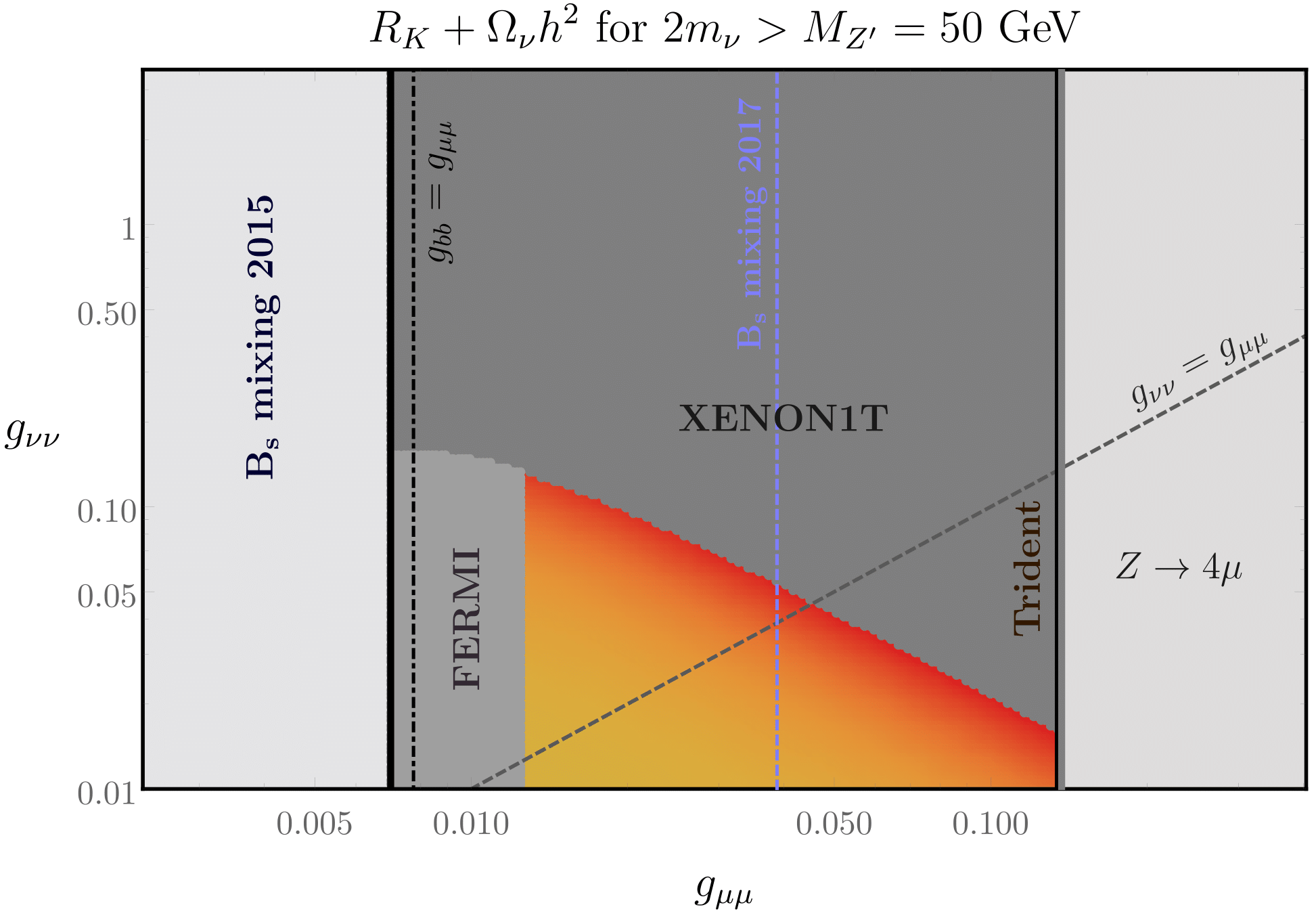}
\end{minipage}
\hfill
\begin{minipage}[t!]{0.49\textwidth}
\includegraphics[height=5.3cm]{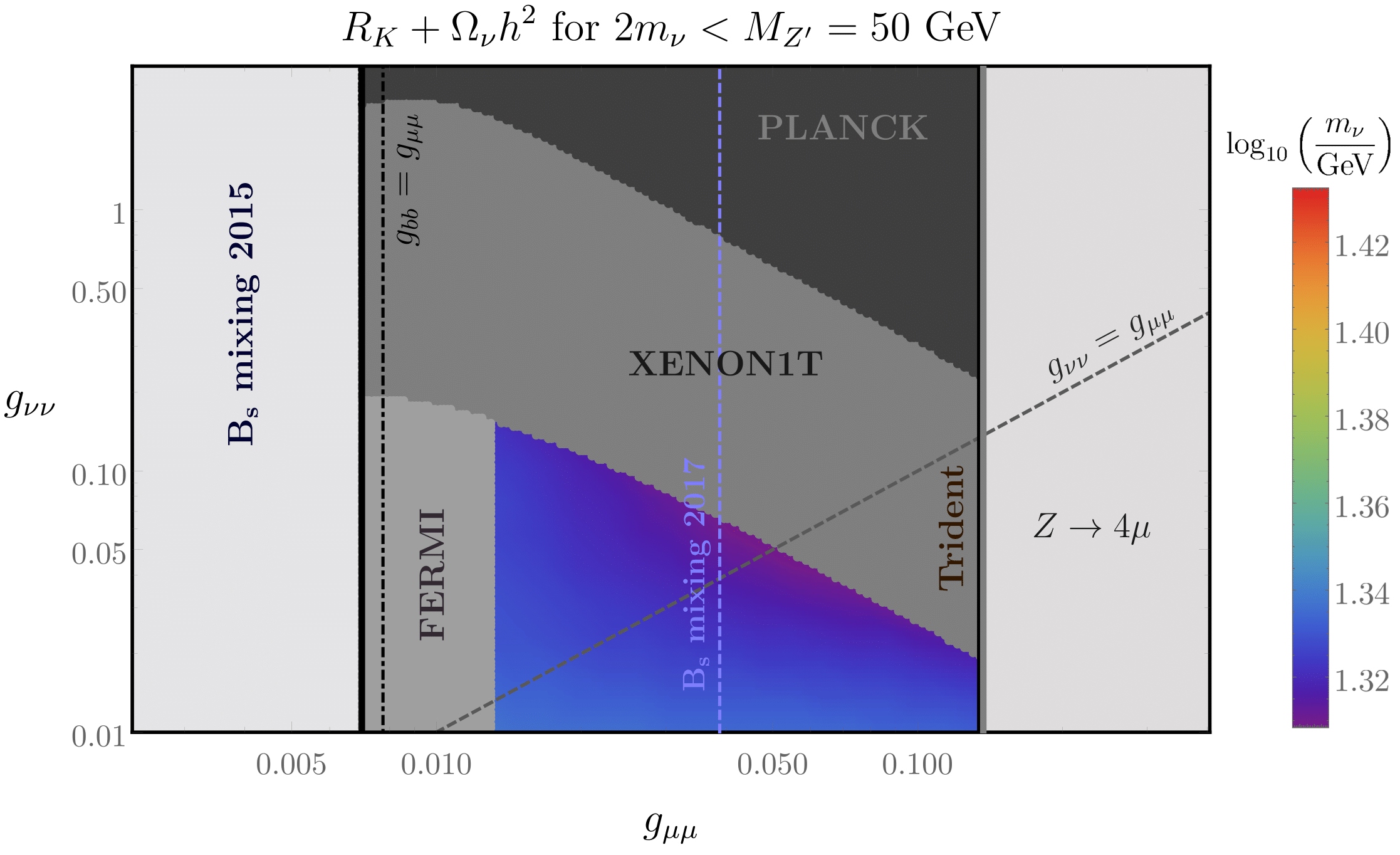}
\end{minipage}
\caption{
\label{fig:mzp50}
Summary of the constraints for $M_{Z^\prime}=50$ GeV.
See text in Section~\ref{conclusion} for details. }
\end{figure}

\begin{figure}
\begin{minipage}[t!]{0.49\textwidth}
\includegraphics[height=5.3cm]{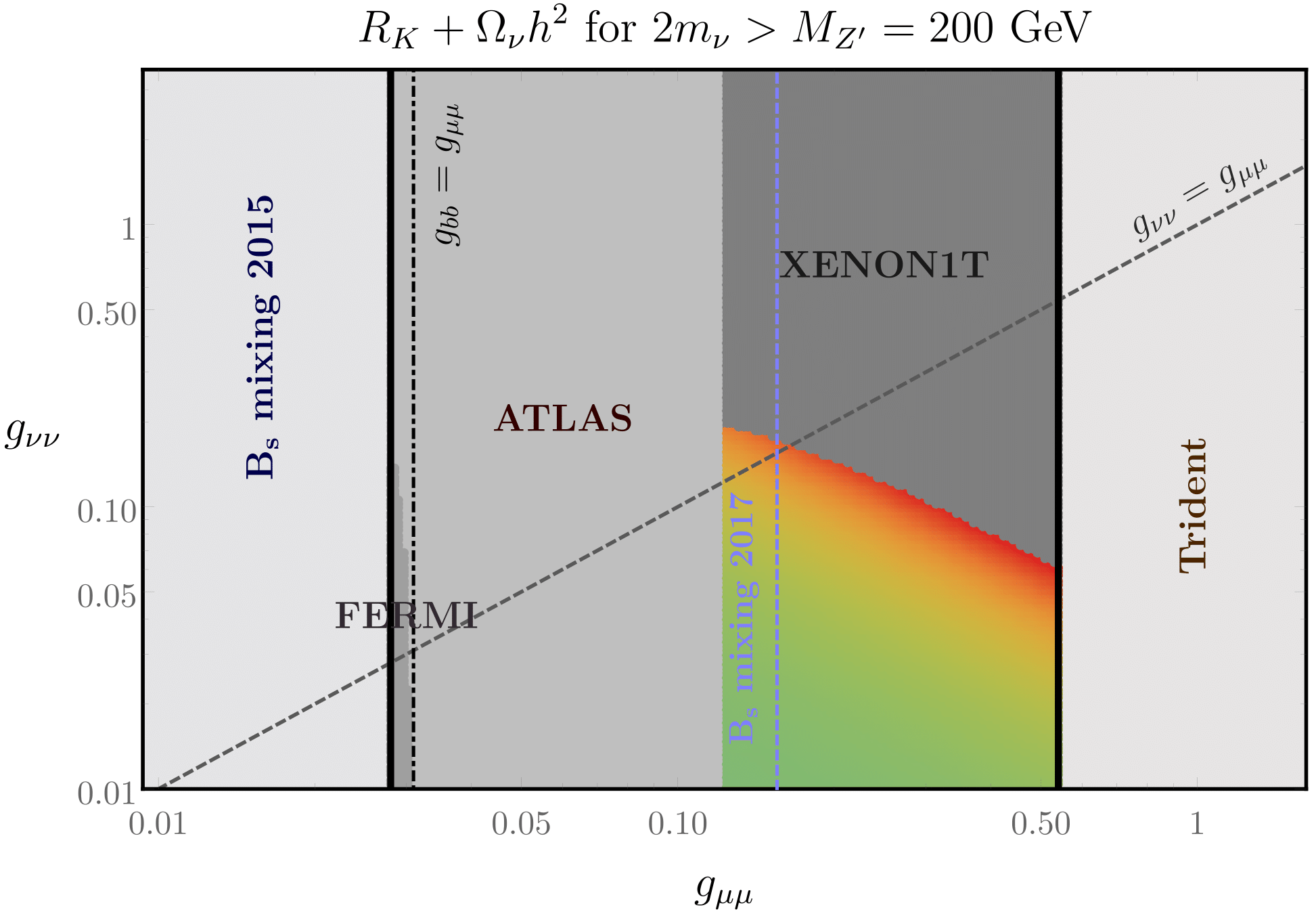}
\end{minipage}
\hfill
\begin{minipage}[t!]{0.49\textwidth}
\includegraphics[height=5.3cm]{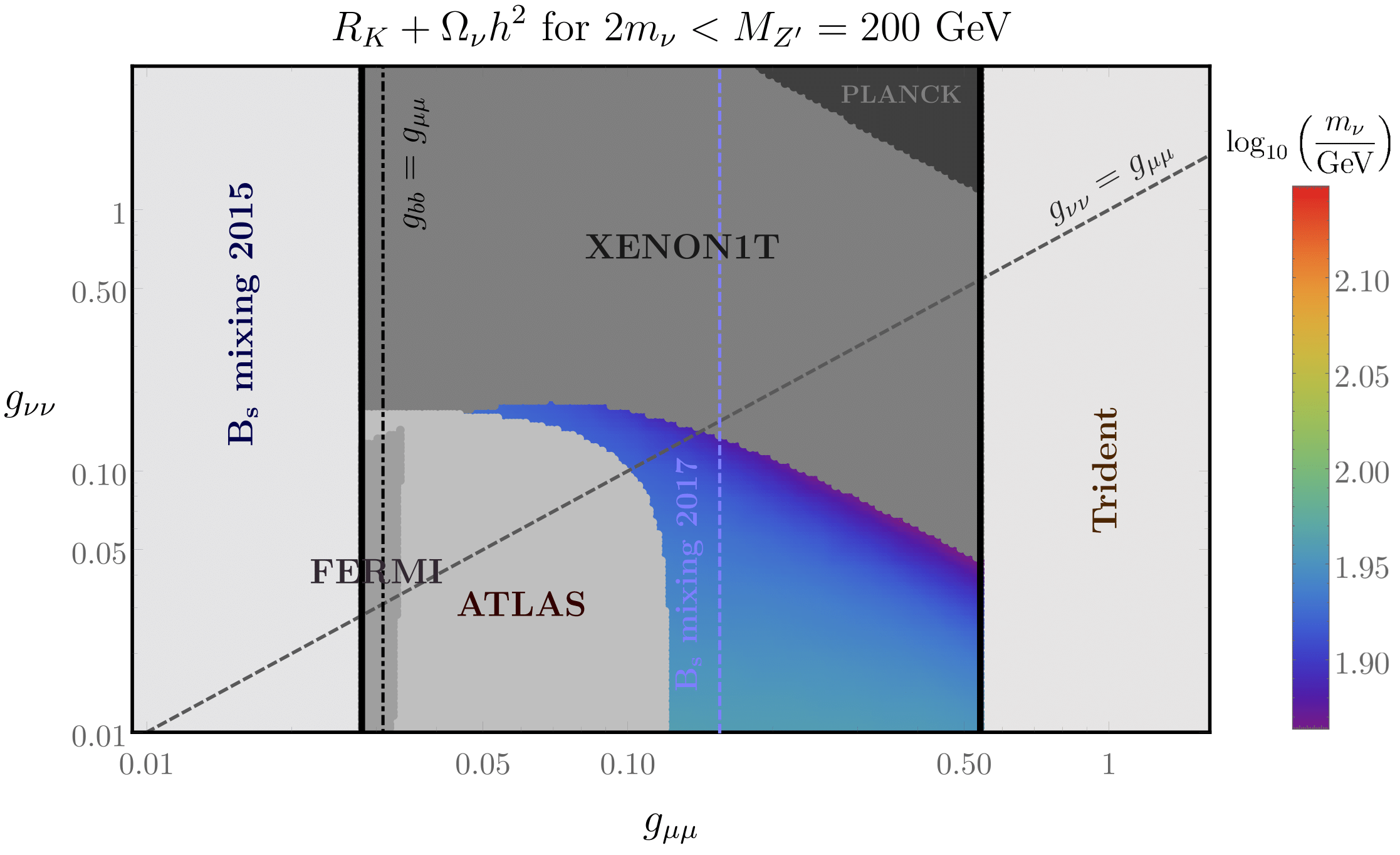}
\end{minipage}
\caption{Summary of the constraints for $M_{Z^\prime}=200$ GeV. See text in Section~\ref{conclusion} for details.}
\end{figure}

\begin{figure}
\begin{minipage}[t!]{0.49\textwidth}
\includegraphics[height=5.3cm]{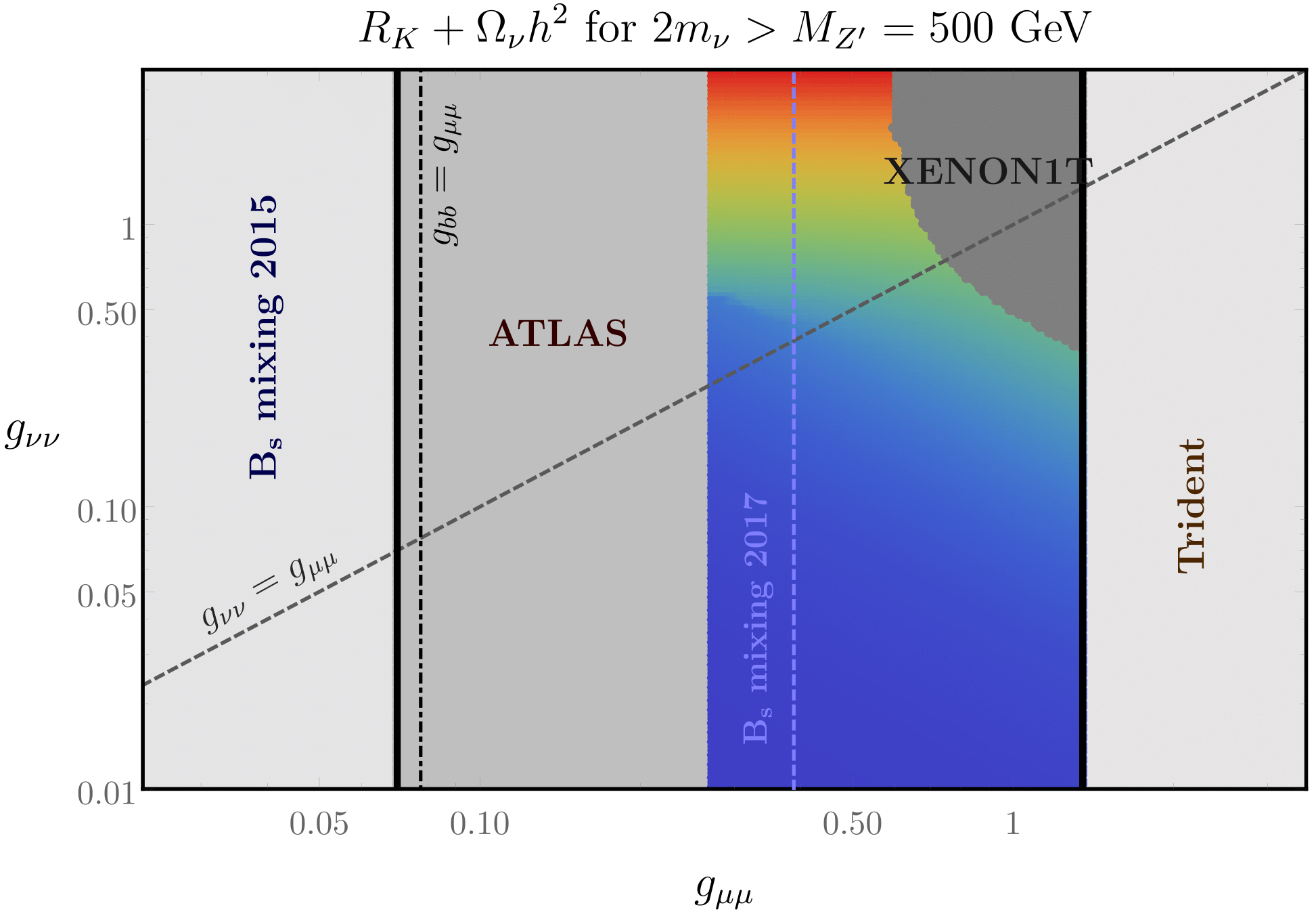}
\end{minipage}
\hfill
\begin{minipage}[t!]{0.49\textwidth}
\includegraphics[height=5.3cm]{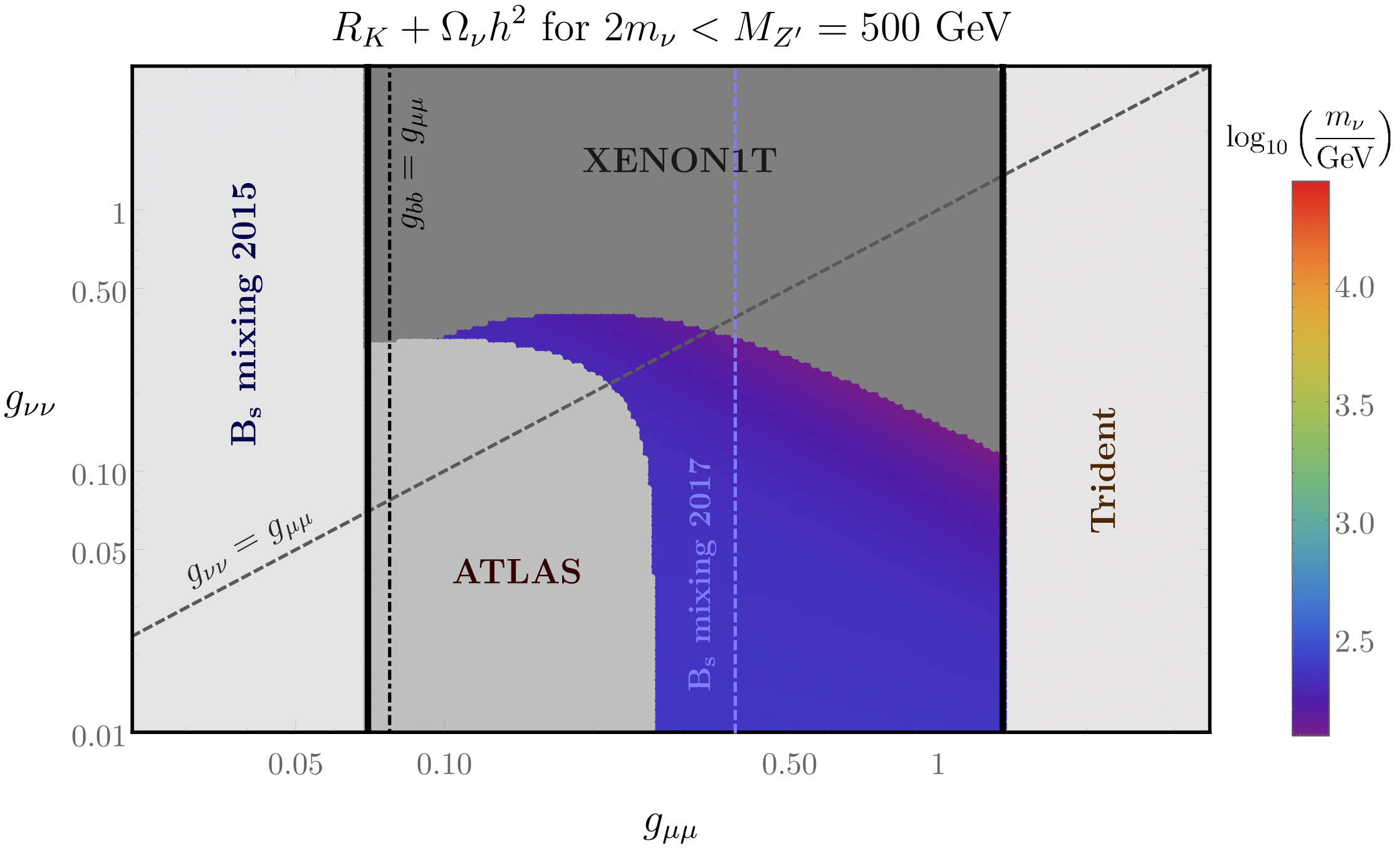}
\end{minipage}
\caption{Summary of the constraints for $M_{Z^\prime}=500$ GeV. See text in Section~\ref{conclusion} for details.}
\end{figure}

\begin{figure}
\begin{minipage}[t!]{0.49\textwidth}
\includegraphics[height=5.3cm]{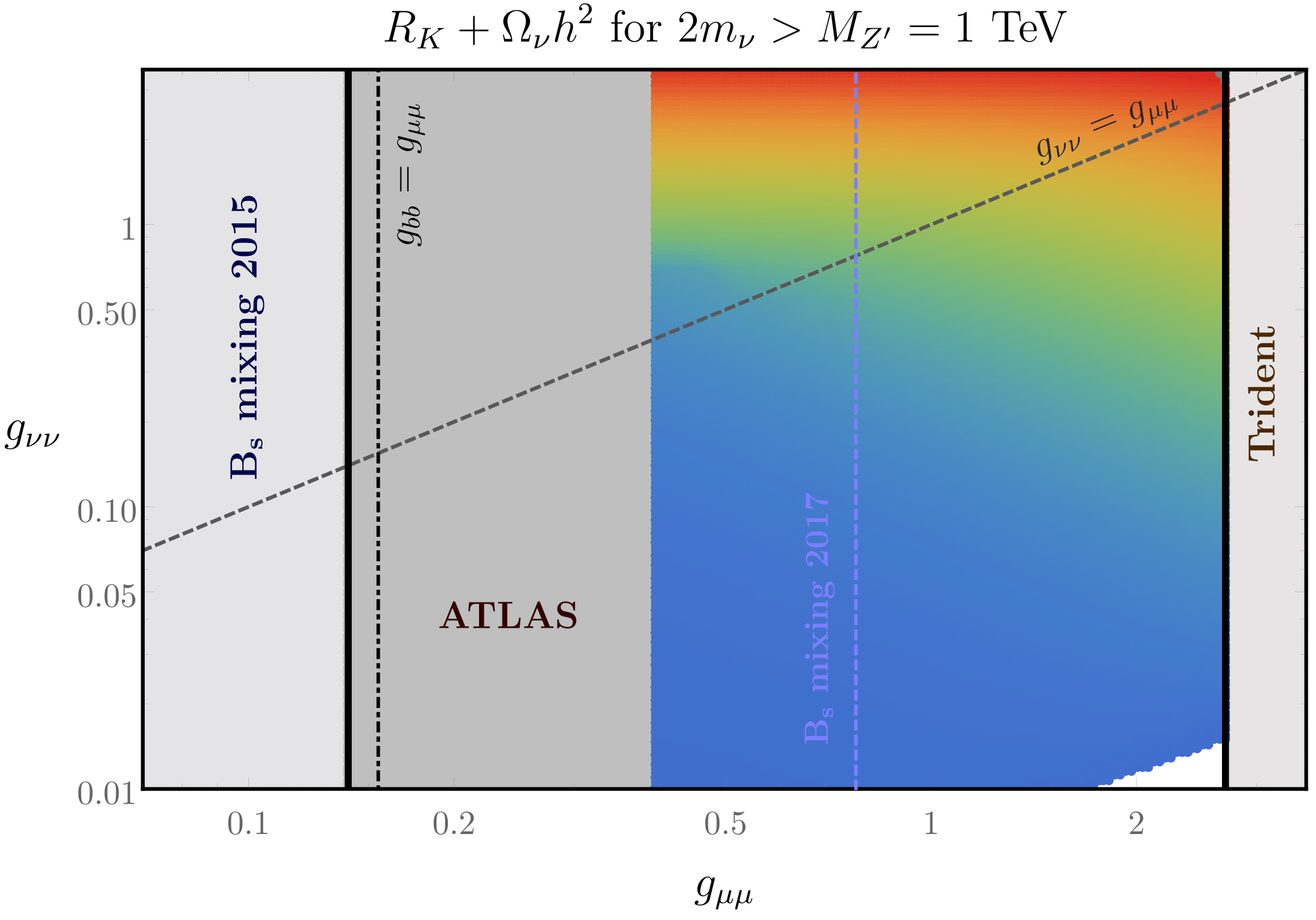}
\end{minipage}
\hfill
\begin{minipage}[t!]{0.49\textwidth}
\includegraphics[height=5.3cm]{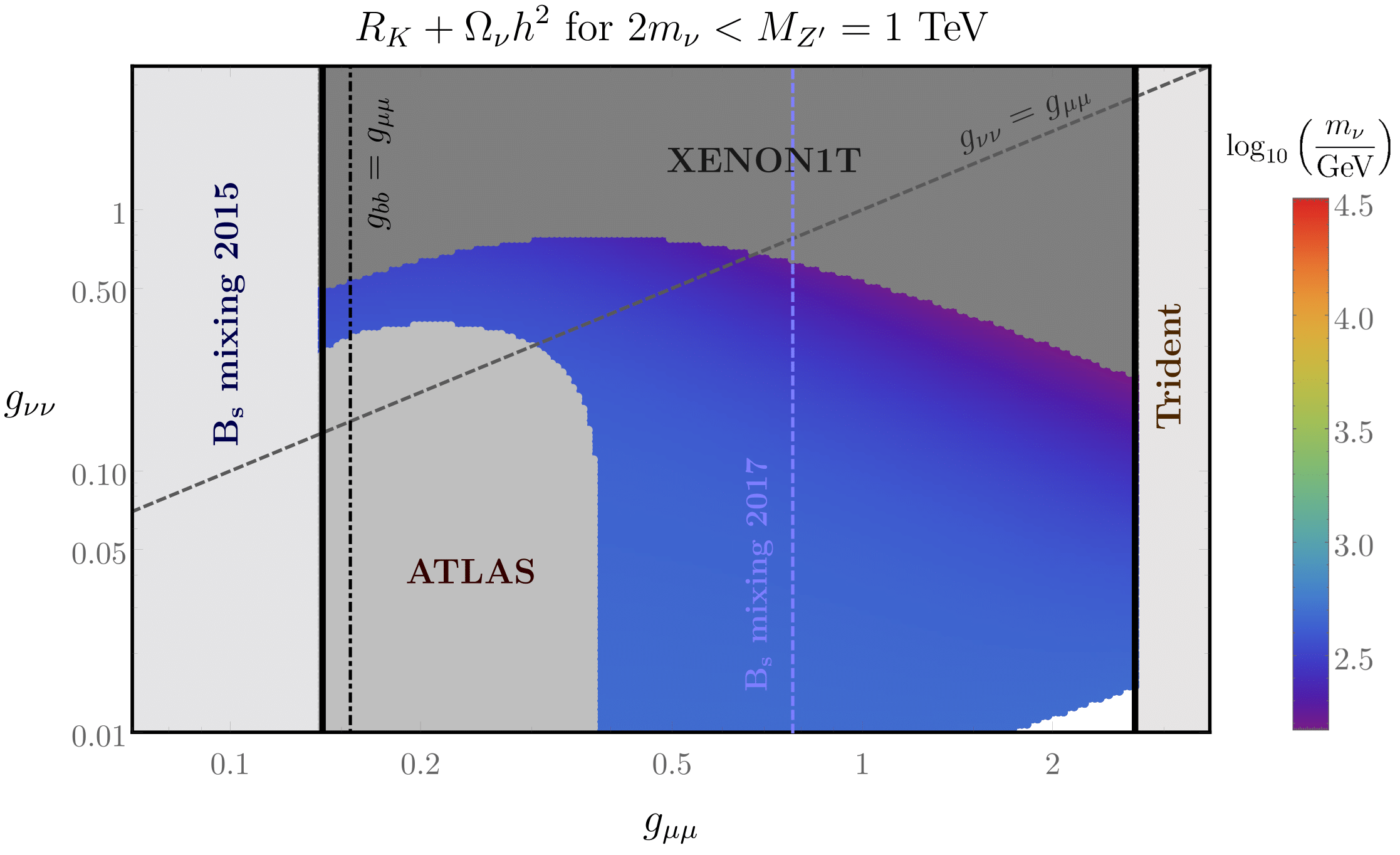}
\end{minipage}
\caption{Summary of the constraints for $M_{Z^\prime}=1$~TeV. See text in Section~\ref{conclusion} for details. }
\end{figure}

\begin{figure}
\begin{minipage}[t!]{0.49\textwidth}
\includegraphics[height=5.3cm]{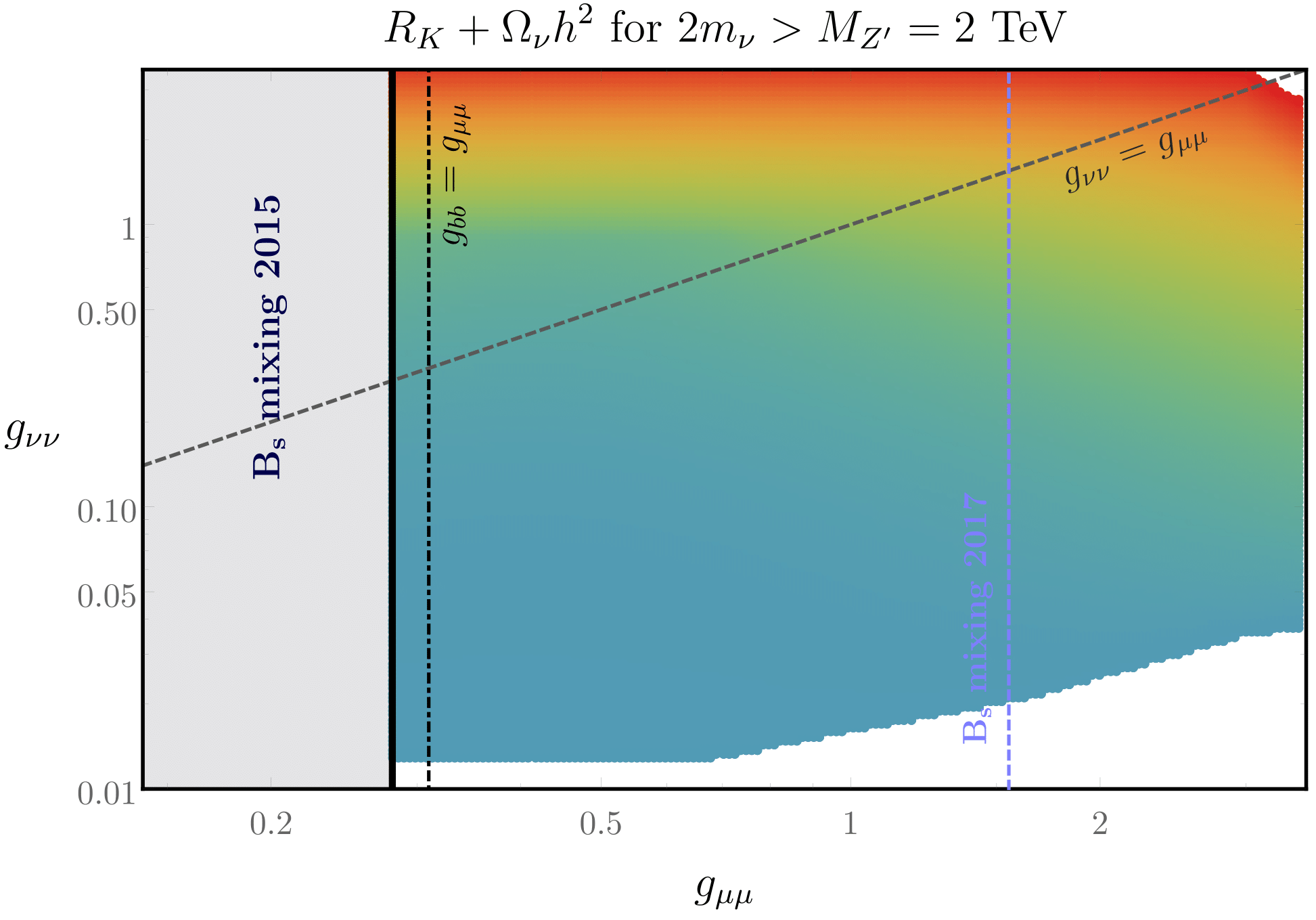}
\end{minipage}
\hfill
\begin{minipage}[t!]{0.49\textwidth}
\includegraphics[height=5.3cm]{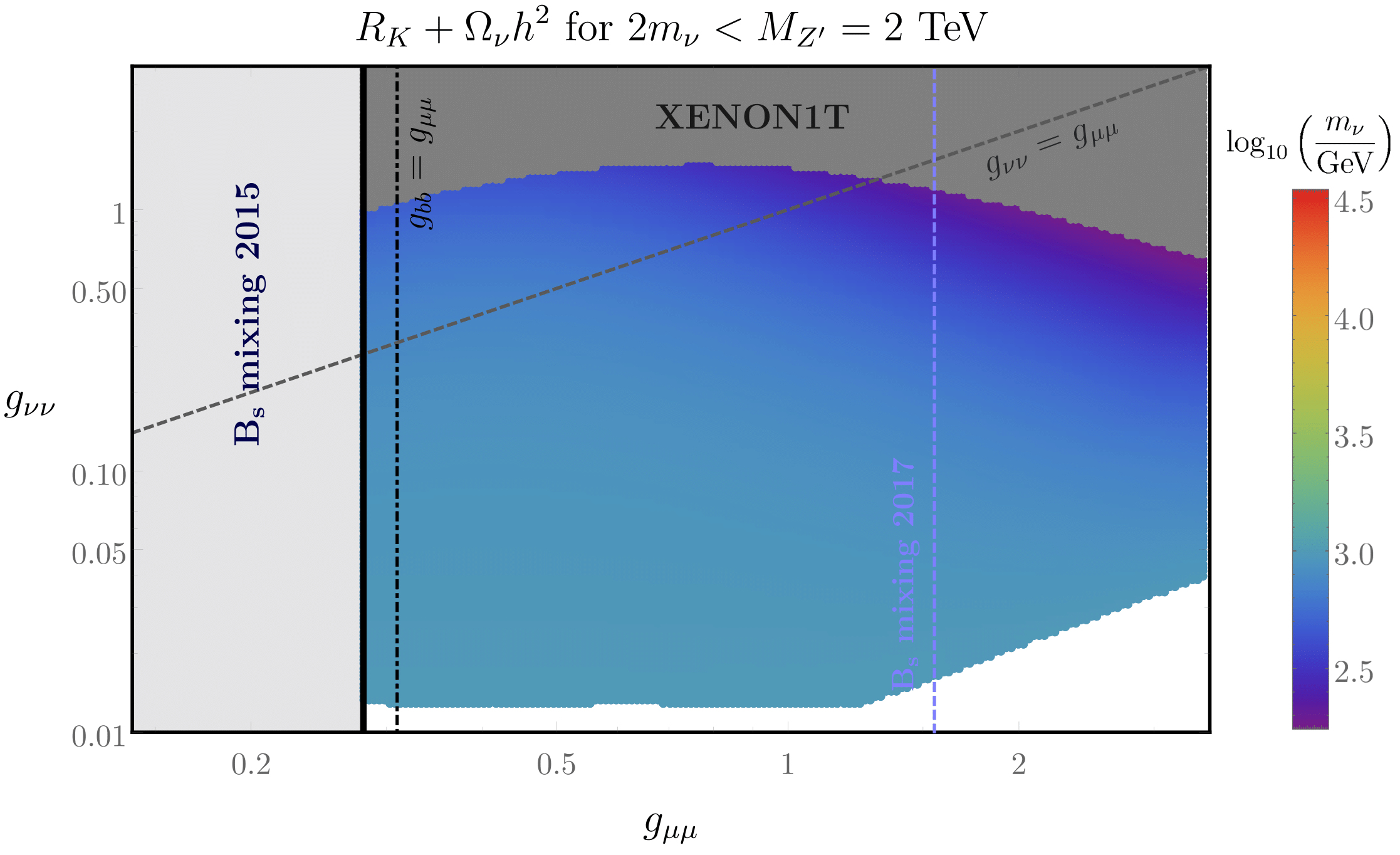}
\end{minipage}
\caption{
\label{fig:mzp2000}
Summary of the constraints for $M_{Z^\prime}=2$~TeV. See text in Section~\ref{conclusion} for details. }
\end{figure}

Our main results are shown in Figs.~\ref{fig:mzp50}-\ref{fig:mzp2000} which show for which parameters  our model can address the B-meson  anomalies while satisfying all experimental and cosmological constraints. 
As discussed below Eq.~\ref{eq:Zp_Rk_couplings}, the relevant parameter space is effectively five-dimensional, and spanned by the $Z^\prime$ couplings to dark matter ($g_{\nu \nu}$), muons ($g_{\mu \mu}$), and $b$ quarks ($g_{bb}$), and by the masses of dark matter ($m_{\nu}$) and the $Z^\prime$ vector messenger ($M_{Z'}$). 
We display it in the $\{g_{\mu \mu}$,$g_{\nu \nu}\}$ plane for several representative values of $M_{Z'}$. 
For each  $g_{\mu \mu}$ and $M_{Z'}$, $g_{bb}$ is fixed according to Eq.~\ref{eq:coeff_RK} to the best fit value reproducing the $R_{K^{(*)}}$ measurements. 
Then $m_{\nu}$ is fixed by the requirement of reproducing the correct relic abundance of dark matter.
There are typically two distinct solutions for $m_{\nu}$ satisfying 
$\langle\sigma v \rangle = \langle\sigma v \rangle_{\rm thermal}$, therefore for each $M_{Z'}$ in the left (right) panel we display the solutions with $m_{\nu} > M_{Z^\prime}/2$ ($m_{\nu} < M_{Z^\prime}/2$).
These solutions are color coded in Figs.~\ref{fig:mzp50}-\ref{fig:mzp2000}, from smaller (blue) to larger (red) $m_{\nu}$. 
The white regions are where we find no parameters choice to tune the annihilation cross section to the thermal value. 
The continuously gray-shaded regions are excluded by direct detection, indirect detection, $B_s$ mixing, dimuon searches at the LHC, $Z$ decay to four muons and/or muon trident constraints.
However, we choose not to shade the region excluded by the recent update of the $B_s$ mixing constraints in \cite{DiLuzio:2017fdq}, and instead represent those by a dashed blue line labeled ``$B_s$ mixing 2017''. The region represented on the left of this line is excluded by these constraints.
For any value of the $Z^\prime$ mass in the considered range there exists a range of parameters reproducing the $R_{K^{(*)}}$ anomalies and the relic abundance, and passing all experimental constraints to date. 
However, for lower $M_{Z^\prime}$ the allowed region corresponds to $g_{\nu \nu} \lesssim g_{\mu \mu}$, once the direct (XENON1T) and indirect (Planck) detection constraints together with updated $B_s$ mixing constraints are taken into account. 
In our model the $Z^\prime$ coupling to muons is suppressed by a mixing angle between the SM 2nd generation lepton doublet and the 4th generation vector-like lepton doublet, and thus we expect  $g_{\nu \nu} \gg g_{\mu \mu}$. 
Conversely, $g_{\nu \nu} \lesssim g_{\mu \mu}$ is unnatural and would require a large hierarchy between the corresponding $U(1)^\prime$ charges, $q_{\nu_4} \ll q_{L_4}$. 
On the other hand, for $300~{\rm GeV} \lesssim M_{Z'} \lesssim 1$~TeV we find some allowed parameter space where $g_{\nu \nu}$ is a factor of few larger than $g_{\mu \mu}$, which is plausible.  
Further increasing $M_{Z'}$ requires a sizable $Z^\prime$ coupling to muons in order to address the B-meson anomalies, $g_{\mu \mu} \gtrsim 1$. 
Then we are forced back into the unnatural $g_{\nu \nu} \sim g_{\mu \mu}$ region, simply  due to perturbativity constraints on $g_{\nu \nu}$ rather than some experimental bounds.   

To summarize, assuming our model is indeed the correct explanation of the observed $R_{K^{(*)}}$ anomalies and dark matter relic abundance, our analysis hints at a particular corner of the parameter space where $300~{\rm GeV} \lesssim M_{Z'} \lesssim 1$~TeV, $m_{\nu} \gtrsim 1$~TeV, $g_{\nu \nu} \gtrsim 1$, $g_{bb} \sim 0.1 g_{\mu \mu}$ and $0.1 \lesssim g_{\mu \mu} \lesssim 1$. 
This viable space implies large mixing with the vector-like fermions to avoid the gauge coupling $g'$ getting into the non-perturbative limit, since $g_{\mu \mu}=g'q_{L_4}(s^L_{24})^2$. The mixing angles are proportional to the VEVs of the scalar fields, $\left<\phi_{\psi}\right>$, while inversely proportional to the mass of the vector-like fermions. Furthermore, the mass of the $Z'$ is generated by the VEVs of the scalar fields, so that $M_{Z'}\sim g'\left<\phi_{\psi}\right>$, which sets the scale of the $U(1)'$ breaking not far from the TeV scale, $\left<\phi_{\psi}\right>\sim$ TeV. This set an upper limit in the vector-like fermions at around this scale to get the necessary large mixing. This limit is far from the current heavy charged mass bounds which sets $M^L_4 \gtrsim 100$ GeV \cite{Patrignani:2016xqp,Poh:2017tfo,Dermisek:2013gta}.

\begin{figure}
\begin{minipage}[t!]{0.49\textwidth}
\includegraphics[height=5.3cm]{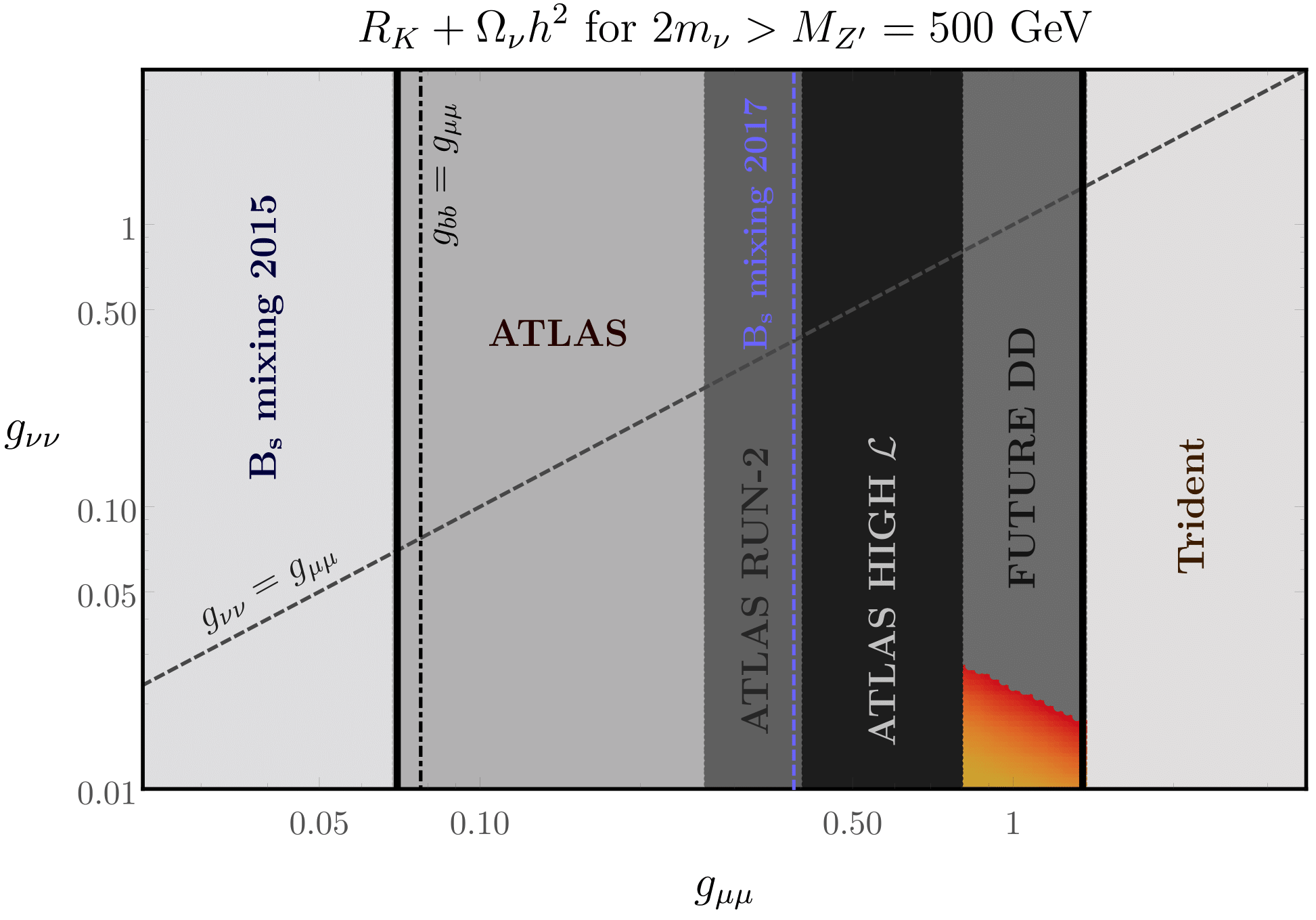}
\end{minipage}
\hfill
\begin{minipage}[t!]{0.49\textwidth}
\includegraphics[height=5.3cm]{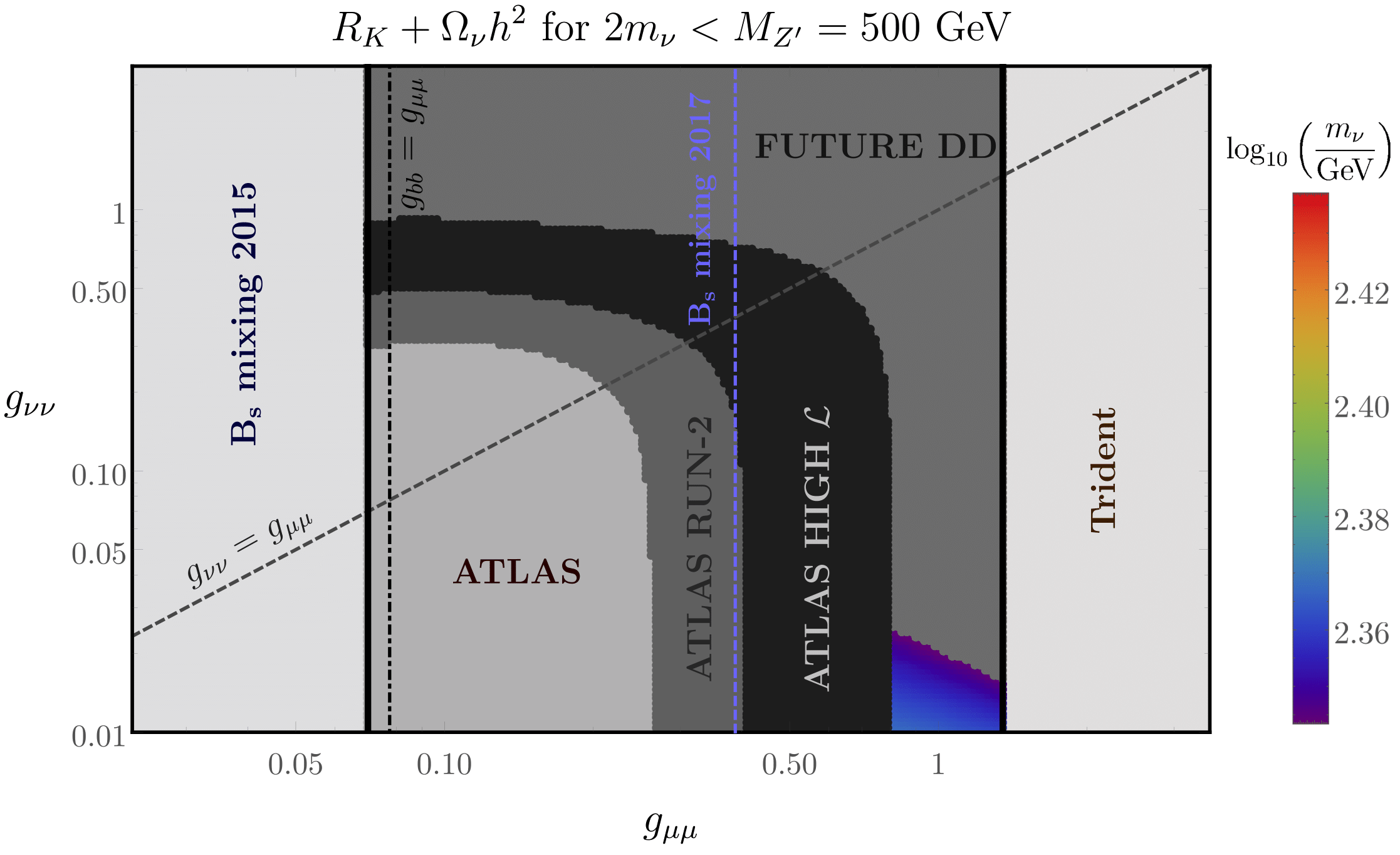}
\end{minipage}
\caption{
\label{fig:mzp500prospects}
Projection of future constraints on the parameter space of our model for $M_{Z^\prime}=500$~GeV. 
The current ATLAS dimuon limits \cite{Aaboud:2017buh} are scaled with integrated luminosity to ${\cal L} = 200$~fb${}^{-1}$ (ATLAS RUN-2) and ${\cal L} = 3000$~fb${}^{-1}$(ATLAS HIGH ${\cal L}$).
Future direct detection limits (FUTURE DD) assume that the current XENON1T~\cite{Aprile:2017iyp} constraints on the DM-nucleon scattering cross section are improved by two orders of magnitude, which roughly corresponds to the projected sensitivity of LZ~\cite{Akerib:2018lyp}, DARWIN~\cite{Aalbers:2016jon}, and Darkside-20k~\cite{Aalseth:2017fik} experiments.   
}
\end{figure}

Incidentally, that parameter space can be probed by several distinct methods. 
First of all, the allowed  window can be further squeezed by better precision measurements of the trident $\nu_\mu N \to \mu^+ \mu^- \nu_\mu N$ process, and by improving the theoretical precision of the SM prediction for the $B_s$ meson mass difference. 
The above statement is in fact valid for all models where  the B-anomalies are addressed by a tree-level $Z'$ exchange.
What is more specific to models where the  $Z'$ interactions with the SM fermions originates from mixing of the latter with vector-like fermions is a non-vanishing $Z'$ coupling not only to muons but also to b-quarks.   
This results in a non-negligible rate of the  partonic process $b \bar b \to Z' \to \mu^+ \mu^-$ which  can be probed by dimuon resonance searches at the LHC. 
In fact, the preferred $M_{Z'}$ range is where the LHC sensitivity is optimal. 
Targeted searches for b-quark-collision initiated process (rather than recast of generic dimuon searches) could lead to a discovery signal in the near future, or to better constraints that are more stringent than the $B_s$ mixing one. 
Finally, the preferred range of dark matter masses and couplings can be probed by direct detection experiments, such that the improvements of one or two orders of magnitude in sensitivity in the next years, which is expected to be achieved by the LZ~\cite{Akerib:2018lyp}, DARWIN~\cite{Aalbers:2016jon} and DarkSide-20k~\cite{Aalseth:2017fik} experiments. 
As illustrated in Fig.~\ref{fig:mzp500prospects}, these future improvements should exclude the remaining most natural parameter space of our model.

\acknowledgments
The authors would like to thank  Debtosh Chowdhury, Darius Faroughy, Yann Mambrini, and Olcyr Sumensari for fruitful discussions. S. F. K. acknowledges the STFC Consolidated Grant ST/L000296/1.  This project has received funding from the European Union's Horizon 2020 research and innovation programme under the Marie Sklodowska-Curie grant agreements Elusives ITN No. 674896 and InvisiblesPlus RISE No. 690575.

\bibliography{bibfile}{}
\end{document}